\documentclass[10pt,a4paper,twocolumn]{article}
\usepackage[top=75pt,bottom=90pt,left=60pt,right=58pt]{geometry}
\usepackage[numbers]{natbib}
\usepackage{amssymb}
\usepackage{amsmath}
\usepackage{latexsym}
\usepackage[table,usenames,dvipsnames]{xcolor}
\usepackage{tabularx}
\usepackage{fourier}
\usepackage{upgreek}
\usepackage{longtable}
\usepackage{varwidth}
\usepackage{multirow}
\usepackage{pbox}
\usepackage[boxed]{algorithm2e}
\usepackage{spverbatim}
\usepackage{url}
\usepackage{setspace}
\usepackage{lineno}
\usepackage{booktabs}
\usepackage{multirow}
\usepackage{authblk}
\usepackage{marginnote}
\usepackage[misc]{ifsym}
\usepackage{tikz}

\modulolinenumbers[5]

\DeclareMathAlphabet{\mathpzc}{T1}{pzc}{m}{it}

\usepackage[bookmarks, colorlinks, breaklinks, pdftitle={},pdfauthor={Camille Roth}]{hyperref}
\hypersetup{linkcolor= MidnightBlue,citecolor= MidnightBlue,filecolor=black,urlcolor= MidnightBlue}

\renewcommand\emph[1]{{\textit{#1}}}

\newcommand{\ra}[1]{$\mathbf{\rightarrow}$~#1}

\newcommand{\tb}[1]{\textcolor{blue}{#1}}

\newcommand{\x}[1]{\textbf{\underline{#1}}}

\DeclareMathAlphabet{\mathpzc}{T1}{pzc}{m}{it}

\def\checkmark{\tikz\fill[scale=0.4](0,.35) -- (.25,0) -- (1,.7) -- (.25,.15) -- cycle;}%

\begin{document}

\renewcommand\Affilfont{\fontsize{9}{10.8}\itshape}

\title{Semantic Hypergraphs}
\author[1]{Telmo Menezes\thanks{menezes@cmb.hu-berlin.de (\Letter)}}
\author[1,2]{Camille Roth\thanks{roth@cmb.hu-berlin.de}}

\affil[1]{\href{http://cmb.huma-num.fr}{Computational Social Science Team}\\Centre Marc Bloch Berlin (CNRS/HU), Friedrichstr. 191, 10117 Berlin, Germany}
\affil[2]{CAMS (Centre Analyse et Math\'ematique Sociales, UMR 8557 CNRS/EHESS), 54 Bd Raspail, 75007 Paris, France}

\date{}

\maketitle

\abstract{Approaches to Natural language processing (NLP) may be classified along a double dichotomy open/opaque -- strict/adaptive. The former axis relates to the possibility of inspecting the underlying processing rules, the latter to the use of fixed or adaptive rules. We argue that many techniques fall into either the \emph{open-strict} or \emph{opaque-adaptive} categories. Our contribution takes steps in the \emph{open-adaptive} direction, which we suggest is likely to provide key instruments for interdisciplinary research.
The central idea of our approach is the \emph{Semantic Hypergraph (SH)}, a novel knowledge representation model that is intrinsically recursive and accommodates the natural hierarchical richness of natural language. The SH model is hybrid in two senses. First, it attempts to combine the strengths of ML and symbolic approaches. Second, it is a formal language representation that reduces but tolerates ambiguity and structural variability. 
We will see that SH enables simple yet powerful methods of pattern detection, and features a good compromise for intelligibility both for humans and machines. It also provides a semantically deep starting point (in terms of explicit meaning) for further algorithms to operate and collaborate on. We show how modern NLP ML-based building blocks can be used in combination with a random forest classifier and a simple search tree to parse NL to SH, and that this parser can achieve high precision in a diversity of text categories. We define a pattern language representable in SH itself, and a process to discover knowledge inference rules. We then illustrate the efficiency of the SH framework in a variety of tasks, including conjunction decomposition, open information extraction, concept taxonomy inference and co-reference resolution, and an applied example of claim and conflict analysis in a news corpus.}

\medskip

\newcommand{\sep}{;}

\section{Introduction}
Natural language processing (NLP) approaches generally belong to either one of two main strands, which also often appear to be mutually exclusive. On the one hand we essentially have symbolic methods and models which are open, in the sense that their internal mechanisms as well as their conclusions are easy to inspect and understand, but which deal with linguistic patterns in a relatively strict manner. 
On the other hand, we have adaptive models based on machine learning (ML) which are usually opaque to inspection and too complex for their reasoning to be intelligible, but which achieve increasingly impressive feats that suggest deeper understanding.

Presently, there is a strong research focus on the latter, and for good reason. Among adaptive models, deep neural networks, for one, managed to jointly learn and improve performance in classic NLP tasks such as part-of-speech tagging, chunking, named-entity recognition, and semantic role-labeling~\cite[as early as][]{collobert2008unified}. 
In other cases, modern ML enabled methods that did not exist before e.g., estimation of semantic similarity using word embeddings~\cite{mikolov2013efficient}. More recently, Bidirectional Encoder Representations from Transformers (BERT) have shown that pre-trained general models can be fine-tuned to achieve state-of-the-art performance in specific language understanding tasks such as question answering and language inference~\cite{devlin-etal-2019-bert}. Nonetheless, symbolic methods possess several proper and important features, namely that they can offer human-readable knowledge representations of knowledge, as well as language understanding through formal and inspectable rule-based logical inference.

Why do we observe this apparent trade-off between openness and adaptivity? Initial approaches to NLP were of a symbolic nature, based on rules written by hand, or in algorithms akin to the ones that are used for programming language interpreters and compilers, such as recursive descent parsers. It became apparent that the diversity of grammatical constructs that can be found in natural language is too large to be tackled in such a way. The problem is compounded by the frequent use of ungrammatical constructs that are nevertheless frequent in real-world language usage (e.g. simple mistakes, neologisms, slang). In other words, content in natural language is generated by actors that are much more complex, and also more error-prone and error-tolerant than conventional algorithms. ML is a natural fit for this type of problem and, as we mentioned, vastly surpasses the capabilities of human-created symbolic systems in a variety of tasks. 

We suggest however that there is a ``hidden relationship'' between explicit symbolic manipulation rules and modern ML: the latter can be seen as a form of ``automatic programming'' through large-scale statistical learning processes, that amount to the generation of highly complex programs through adaptive pressure instead of human programmers' efforts. It does not matter if it is gradient descent on a multi-layered network topology, or something more prosaic like entropy reduction in a decision tree, it is still program generation through adaptation. The capability of these methods to generate such complex programs is what allows them to tackle the complexities of NL, but it is also this very complexity that makes them opaque.

We can thus imagine a double dichotomy open/opaque -- strict/adaptive. We argue that existing approaches generally fall into either the open-strict or \emph{opaque-adaptive} categories. 
A few approaches have ventured into the \emph{open-adaptive} domain~\cite{banarescu2013abstract,mausam2012open} and our contribution aims at significantly expanding this direction.
Before discussing our approach, let us consider why \emph{open-adaptive} is a desirable goal. 
The work we present here was performed in the context of a computational social science (CSS) research team, where NLP is a scientific instrument capable of assisting in the analysis of text corpora that are too vast for humans to study in detail.
We argue that further progress in the study of socio-technical systems and their dynamics could be enabled by \emph{open-adaptive} scientific instruments for language understanding.

In current CSS research, the more common approaches aim to transform natural language documents into structured data that can be more easily analyzed by scholars and are referred to by a variety of umbrella terms such as ``{text mining}''~\cite{srivastava2009text}, ``{automated text analysis}''~\cite{grimmer-stewart-2013} or ``{text-as-data methods}''~\cite{wilkerson2017large}. They exhibit a wide range of sophistication, from simple numerical statistics to more elaborate ML algorithms. Some methods indeed rely essentially on scalar numbers, for instance by measuring text similarity (e.g., with cosine distance~\cite{singhal2001modern}) or attributing a valence to text, as in the case of ideological estimation~\cite{sim-2013-mea} or sentiment analysis~\cite{pang2008opinion}, which in practice may be used to appraise the emotional content of a text (anger, happiness, disagreement, etc.) or public sentiment towards political candidates in social media~\cite{wang2012system}. Similarly, political positions in documents may be inferred from so-called ``Wordscores''~\cite{lowe2008understanding} -- a popular method in political science that also relies on the summation of pre-computed scores for individual words, and has more refined elaborations, e.g. with Bayesian regularization~\cite{monroe2008fightin}. Other methods preserve the level of words: such is the case with term and pattern extraction (i.e., discovering salient words through the use of helper measures like term frequency--inverse document frequency (TF-IDF)~\cite{salton1988term}), so-called ``Named Entity Recognition''~\cite{nadeau2007survey} (used to identify people, locations, organizations, and other entities mentioned in corpuses, for example in news corpora~\cite{diesner-2005-revealing} or Twitter streams~\cite{ritter2011named}) and ad-hoc uses of conventional computer science approaches such as regular expressions to identify chunks of text matching against a certain pattern (for example, extracting all p-values from a collection of scientific articles~\cite{chavalarias2016evolution}). Another strand of approaches operates at the level of word sets, including those geared at topic detection (such as co-word analysis~\cite{leydesdorff-2017-co-word}, Latent Dirichlet Allocation (LDA)~\cite{blei2003latent} and TextRank~\cite{mihalcea2004textrank}, used to extract the topics addressed in a text) or used for relationship extraction (meant at deriving semantic relations between entities mentioned in a text, e.g., is(Berlin, City)) \cite{angeli2015leveraging}. Recent advances in embedding techniques have also made it possible to describe topics extensionally as clusters of documents in some properly defined space \cite{angelov2020top2vec,le2014distributed}.

Overall, these techniques provide useful approaches to analyze text corpora at a high level, for example, with regard to their main entities, relationships, sentiment, and topics. However, there is limited support to detect, for instance, more sophisticated claim patterns across a large volume of texts, what recurring statements are made about actors or actions, and what are the qualitative relationships among actors and concepts. This type of goal, for example, extends semantic analysis to a socio-semantic framework \citep{roth-sociacs} which also takes into account actors who make claims or who are the target of claims \cite{diesner-2005-revealing}.

It is also particularly interesting to consider the model of knowledge representation that is implicitly or explicitly associated with the various NLP/text mining/information extraction approaches. To illustrate, on one extreme we can consider traditional knowledge bases and semantic graphs, which are open in our sense, but also limited in their expressiveness and depth. On the other, we have the extensive knowledge opaquely encoded in neural network models such as BERT or GPT-2/3~\cite[e.g.][]{brown2020language}. Beyond the desirability of open knowledge bases for their own sake, we propose that a language representation that is convenient for both humans and machines can constitute a \emph{lingua franca}, through which systems of cognitive agents of different natures can cooperate in a way that is understandable and inspectable. Such systems could be used 
to combine the strengths of symbolic and statistical inference.

The central idea of our approach is the \emph{Semantic Hypergraph (SH)}, a novel knowledge representation model that is intrinsically recursive and accommodates the natural hierarchical richness of NL. The SH model is hybrid in two senses. First, it attempts to combine the strengths of ML and symbolic approaches. Second, it is a formal language representation that reduces but tolerates ambiguity, and that also reduces structural variability. We will see that SH enables simple methods of pattern detection to be more powerful and less brittle, that it is a good compromise for intelligibility both for humans and machines, and that it provides a semantically deeper starting point (in terms of explicit meaning) for further algorithms to operate and collaborate on.

In the next section we discuss the state of the art, comparing SH to a number of approaches from various fields and eras. We then describe the structure and syntax of SH, followed by an explanation on how modern and standard NLP ML-based building blocks provided by an open source software library~\cite{honnibal2015improved} can be used in combination with a random forest classifier and a simple search tree to parse NL to SH. Here we also provide precision benchmarks of our current parser, which is then employed in the experiments that follow. We attempted to perform a set of experiments of a rather diverse nature, to gather evidence of SH usefulness in a variety of roles, and of its potential to tackle the challenge that we started by stating in this introduction, and to gather empirical insights. One important language understanding task is information extraction from text. One formulation of such a task that attracts significant attention is that of Open Information Extraction (OIE) --- the domain-free extraction from text of tuples (typically triplets) representing semantic relationships~\cite{etzioni2008open}. We will show that a small and simple set of SH patterns can produce competitive results in an OIE benchmark, when pitted against more complex and specialized systems in that domain. We will demonstrate concept taxonomy inference and co-reference resolution, followed by claim and conflict identification in a database of news headers. We will show how SH can be used to generate semantically rich visual summaries of text.

\section{Related Work}


\paragraph{Knowledge bases.}
As a knowledge representation formalism, it is interesting to compare SH with traditional approaches. Let us start with 
triplet-based ones. For example, the \emph{Semantic Web}~\cite{berners2001publishing, shadbolt2006semantic} community tends to use standards such as \emph{RDFa}~\cite{adida2008rdfa}, which represent knowledge as \emph{subject-predicate-object} expressions, and are conceptually equivalent to semantic graphs~\cite{allen1982s, sowa2014principles} (similarly, a particular type of hypergraph has been used in~\cite{cattuto2007network} to represent tagged resources by users, yet this also reduces to fixed triplet conceptualization). Despite their usefulness for simple cases, such approaches cannot hope to match the semantic sophistication of what can be conveyed with open text. Binary relationships and lack of recursion limit the expressive power of semantic graphs, and we sill see how SHs can represent semantic information that is lost in the graphic representation, for example the ability to express $n$-ary relationships, propositions about propositions and constructive definitions of concepts.

A further type of approaches relying on knowledge bases is epitomized by the famous Cyc~\cite{lenat1990cyc} project, a multi-decade enterprise to build a general-purpose and comprehensive system of concepts and rules. It is an impressive effort, nevertheless hindered by the limitations that we alluded to in the previous section concerning the ambiguity and diversity of semantic structures contained in NL, given that it relies purely on symbolic reasoning. Cyc belongs to a category of systems that are mostly concerned with question answering, a different aim that the one of the work that we propose here, which is more concerned with aiding in the analysis and summarization of large corpora of text for research purposes, especially in the social sciences, while not requiring full disambiguation of meaning nor perfect reasoning or understanding.

Several other notable knowledge bases of a similar semantic graph nature have been developed, some relying on collaborative human efforts to gather ground assertions, for example MIT's ConceptNet~\cite{speer2017conceptnet}, ATOMIC~\cite{sap2019atomic} from the Allen Institute, or very rigorous scholarly efforts of annotation, as is the case with WordNet~\cite{miller1995wordnet} and its multiple variants, or relying on wiki-like platforms such as WikiData~\cite{vrandevcic2012wikidata}, or mining relationship from Wikipedia proper, as is the case with DBPedia~\cite{auer2007dbpedia}, and more recently a transformer language model has been proposed to automatically extend common-sense knowledge bases~\cite{bosselut2019comet}. We envision that such general-knowledge bases could be fruitfully integrated with SHs for various purposes, but such endeavours are beyond the scope of this work. We are instead interested in demonstrating what can be achieve by going beyond such non-hypergraphic appraches.

\paragraph{Hypergraphic approaches to knowledge representation.}
Hypergraphs have been proposed already in the 1970s as a general solution for knowledge representation~\cite{boley1977directed}. More recently, Ben Goertzel produced similar insights~\cite{goertzel2006patterns}, and in fact included an hypergraphic database called AtomSpace as the core knowledge representation of his OpenCog framework~\cite{hart2008opencog}, an attempt to make Artificial General Intelligence emerge from the interaction of a collection of heterogeneous system. As is the case with Cyc, the goals of OpenCog are however quite distinct from the aim of our work.

A model that shares similarities with ours but purely aims at solving a meaning matching problem is that of \emph{Abstract Meaning Representation (AMR)}~\cite{banarescu2013abstract}. AMR is based on PropBank verbal propositions and their arguments~\cite{palmer2005proposition}, ensuring that all such meaning structures can be represented. SH completeness is based instead on Universal Dependencies \cite{nivre2016universal}, ensuring instead that all cataloged grammatical constructs can be represented. AMR's goal is to purely abstract meaning, while SH accommodates the ambiguity of the original NL utterances, bringing several important benefits: it makes their computational processing tractable in further ways, tolerates mistakes better and preserves communication nuance that would otherwise be lost. Furthermore, it remains open to structures that may not be currently envisioned. Parsing AMR to NL is a particularly hard task and, to our knowledge, there is currently no parser that approaches the capabilities of what we will demonstrate in this work. In part, this is a practical problem: we will see how we can take advantage of intermediary NLP tasks that are well studied and developed to achieve NL to SH parser. Doing the same for AMR requires the construction of training data by extensive annotation efforts by humans. It could be argued that this is still a preferable goal, no matter how distant, given that AMR removes all ambiguity from statements. Here we point out that this aspect of AMR is also a downside, firstly because it makes all failures of understanding catastrophic (we will see how this is not the case for SH), and secondly because NL is inherently ambiguous. It is often the case that even human beings cannot fully resolve ambiguities, or that an ambiguous statement gains importance later on, with more information. We aim to define SH as a \emph{lingua franca} for the collaboration of human an algorithmic actors of several natures, a less rigid goal than the one embodied by AMR.

\paragraph{Free text parsing.}
A classical NLP task is that of making explicit the grammatical structure of a sentence in the form of a parse tree. A particularly common type of such a tree in current use is the Dependency Parse Tree (DPT), based on dependency grammars. We will see that our own parser takes advantage of DPTs (among other high-level grammatical / linguistic features) as intermediary steps, but it is also interesting to notice that DPTs themselves can be considered as a type of hypergraphic representation of language \tb{\cite{reddy2016transforming}}. In fact, as we will discuss below, they are already employed in various targeted language understanding tasks in a CSS context. 

From the perspective of hypergraphic representation of language, the fundamental difference between DPTs and SHs is that the former aims at expressing the grammatical structure of language, while the latter its semantic structure, in the simplest possible way that enables meaning extraction in a principled 
and predictable way. In contrast to the \emph{ad-hoc} nature of information extraction from DPTs, we will see that SHs structure NL in a way akin to functional computer languages, and allow for example for a generic methodology of extracting patterns. The expressive power of such patterns will be demonstrated in several ways, namely by demonstrating competitive results in a standard Open Information Extraction task. We will see that the type system of SHs (relying on $8$ types) is much simpler than the diversity of grammatical roles contained in a typical set of dependency labels (such as Universal Dependencies), and we will also provide empirical evidence that SHs are not isomorphic to DPTs.


In the realm of OIE, one approach in particular with which our work shares some similarities is that of learning \emph{open pattern templates}~\cite{mausam2012open}. These pattern templates combine at the same symbolic level dependency parse labels and structure, part-of-speech tags, explicit lexical constraints and higher-order inferences (e.g. that some term refers to a \emph{person}), to achieve sophisticated language understanding in the extraction of OIE tuples, being able to extract relations that are not only of a verbal nature, and demonstrating sensitivity to context. The work we will present does not attempt to directly combine diverse linguistic features at the service of a specific language understanding task. Instead, we propose to use such features to aid in the translation of NL into a structured representation, which relies by comparison on a very simple and uniform type system, and from which complex NL understanding tasks become easier, and that is of general applicability to a diversity of such tasks, while remaining fully readable and understandable by humans. Furthermore, it defines a system of knowledge representation in itself, that is directly focused on meaning instead of grammar.

\paragraph{Text mining.}
We have already covered in the previous section the most commonly used 
text mining approaches, while emphasizing the relative lack of sophistication in understanding text meaning. The need for such sophistication 
is all the more pregnant for social sciences. On the one hand, qualitative social science methods of text analysis do not scale to the enormous datasets that are now available. Furthermore, quantitative approaches allow for other types of analysis that are enriching and complementary to qualitative research, yet may simplify extensively the processing in such a way that it hinders their adoption by scholars used to the refinement of qualitative approaches. And the more sophisticated the NLP techniques become, the further they tend to be from being used for large-scale text analysis purposes. Indeed, these systems are fast and accurate enough to form a starting point for more advanced computer-supported analysis in a CSS context, and they enable approaches that are substantially more sophisticated than the text mining state of the art discussed above. Yet, the results of such systems may seem relatively simplistic compared to human-level understanding of natural language. 

The literature already features some works which attempt at going beyond language models based on word distributions (such as bags of words, co-occurrence clusters, or so-called ``topics'') or triplets. For instance, Statement Map~\cite{murakami2010statement} is aimed at mining the various viewpoints expressed around a topic of interest in the web. Here a notion of claim is employed. A statement provided by the user is compared against statements from a corpus of text extracted from various web sources. Text alignment techniques are used to match statements that are likely to refer to the same issue. A machine learning model trained over NLP-annotated chunks of text classifies pairs of claims as ``agreement'', ``conflict'', ``confinement'' and ``evidence''. More broadly, the subfield of argumentation mining~\cite{lippi2016argumentation} also makes extensive use of machine learning and statistical methods to extract portions of text corresponding to claims, arguments and premises. These approaches generally rely on surface linguistic features, there is however an increasing trend of dealing with structured and relational data. Already in 2008, 
\cite{van2008parsing} proposed a system to extract binary semantic relationships from Dutch newspaper articles. A recent work~\cite{ruiz2016more} presents a system aimed at analysing claims in the context of climate negotiations. It leverages dependency parse trees and general ontologies \cite{staab2010handbook} to extract tuples of the form: $\langle\text{actor}, \text{predicate}, \text{negotiation\_point}\rangle$ where the actors are stakeholders (e.g., countries), the predicates express agreement, opposition or neutrality and the negotiation point is identified by chunk of text. Similarly, in another recent work~\cite{van2017clause}, parse trees are used to automatically extract source-subject-predicate clauses in the context of news reporting over the 2008-2009 Gaza war, and used to show differences in citation and framing patterns between U.S. and Chinese sources.

These works help demonstrate the feasibility of using parse trees and other modern NLP techniques to identify viewpoints and extract more structured claims from text. Being a step forward from pure bag-of-words analysis, they still leave out a considerable amount of information contained in natural language texts, namely by relying on topic detection, or on pre-defined categories, or on working purely on source-subject-predicate clauses. We propose to introduce a more sophisticated language model, where all entities participating in a statement are identified, where entities can be described as combinations of other entities, and where statements can be entities themselves, allowing for claims about claims, or even claims about claims about claims. The formal backbone of this model consists of an extended type of hypergraph that is both recursive and directed, thus generalizing semantic graphs and inducing powerful representation capabilities.

\section{Semantic hypergraphs -- structure and syntax}
\label{sec:hypergraphs}

\subsection{Structure}

The SH model is essentially a recursive, ordered hypergraph that makes the structure contained in natural language (NL) explicit. On one hand, NL is recursive, allowing for concepts constructed from other concepts as well as statements about statements, and on the other hand, it can express $n$-ary relationships. We will see how a hypergraphic formalism provides a satisfactory structure for NL constructs.  


While a graph $G = (V,E)$ is based on a vertex set $V$ and an edge set $E\subset V\times V$ describing dyadic connections, a \emph{hypergraph}~\cite{battiston2020networks,berge1984hypergraphs} generalizes such structure by allowing $n$-ary connections. In other words, it can be defined as $H = (V, E)$, where $V$ is again a vertex set yet $E$ is a set of hyperedges $(e_i)_{i\in{1..M}}$ connecting an arbitrary number of vertices. Formally, $e_i = \{v_1, ... v_n\} \in E = \mathcal{P}(V)$. 
We further generalize hypergraphs in two ways: hyperedges may be ordered~\cite{eslahchi2007some} and recursive~\cite{iord-hype}. Ordering entails that the position in which a vertex participates in the hyperedge is relevant (as is the case with directed graphs). Recursivity means that hyperedges can participate as vertices in other hyperedges. 
The corresponding hypergraph may be defined as $H=(V,E)$ where $E\subset\mathcal{E}_V$ the recursive set of all possible hyperedges generated by~$V$:
$\mathcal{E}_V=\left\{(e_i)_{i\in\{1..n\}}\,|\,n\in\mathbb{N},\forall i\in\{1..n\}, e_i \in V \cup \mathcal{E}_V\right\}$.
In this sense, $V$ configures a set of irreducible hyperedges of size one i.e., atomic hyperedges which we also denote as \emph{atoms}, similarly to semantic graphs. From here on, we simply call these recursive ordered hyperedges as ``hyperedges'', or just ``edges'', and we denote the corresponding hypergraph as a ``semantic hypergraph''.

\newcommand{\T}[1]{{\small
{\sf #1}}}
\newcommand{\Q}[1]{\begin{center}\T{#1}\end{center}}
\newcommand{\QC}[1]{\vspace{-.12em}\Q{#1}\vspace{-.12em}}
\renewcommand{\QC}[1]{\Q{#1}}

Let us consider a simple example, based on a set $V$ made of four atoms: the noun ``\T{(berlin)}'', the verb ``\T{(is)}'', the adverb ``\T{(very)}'' and the adjective ``\T{(nice)}''. They may act as building blocks for both hyperedges ``\T{(is berlin nice)}'' and ``\T{(very nice)}''.
These structures can further be nested: the hyperedge ``\T{(is berlin (very nice))}'' represents the sentence ``Berlin is very nice''. It illustrates a basic form of recursivity. 

\newcommand{\h}[1]{\T{(#1)}}

\subsection{Syntax}

In a general sense, the hyperedge is the fundamental unifying construct that carries information within the SH formalism. We further introduce the notion of hyperedge \emph{types}, which simply describe the type of construct that some hyperedge represents: for instance, concepts, predicates or relationships, as in the above examples --- respectively \T{(berlin)}, \T{(is)} and \T{(is berlin nice)}. 
We extensively detail hyperedge types and their role in the next subsections. For now, it is enough to know that predicates, in particular and for instance, belong to a larger family of types that are crucial for the construction of hyperedges and that we call \emph{connectors}.  
In this regard, semantic hypergraphs rely on a syntactic rule that is both simple and universal: the first element in a non-atomic hyperedge must be a connector.

In effect, a hyperedge represents information by combining other (inner) hyperedges that represent information. The purpose of the connector is to specify \emph{in which sense} inner hyperedges are connected. Naturally, it can be followed by one or more hyperedges which play the role of arguments with respect to the connector. As hyperedges, if they are not atoms, they must also start with a connector themselves, in a recursive fashion.

We illustrate this on the hyperedge \h{is berlin (very nice)}: here, \h{is} is a predicate playing the role of connector while \h{berlin} and \h{very nice} are arguments of the initial hyperedge. \h{berlin} is an atomic hyperedge, while \h{very nice} is a hyperedge made of two elements: the connector, \h{very}, an atomic hyperedge, and an argument, \h{nice}, also an atomic hyperedge. Both cannot be decomposed further.

\smallskip

Readers who are familiar with Lisp will likely have noticed that hyperedges are isomorphic to \emph{S-expressions}~\cite{mccarthy1960recursive}. This is not purely accidental. Lisp is very close to $\lambda$-calculus, a formal and minimalist model of computation based on function abstraction and application. The first item of an S-expression specifies a function, the following ones its arguments. One can think of a function as an association between objects. Albeit hyperedges do not specify computations, connectors are similar to functions at a very abstract level, in that they define associations. The concepts of ``race to space'' and ``race in space'' are both associated to the concepts ``race'' and ``space'', but the combination of these two concepts yields different meaning by application of either the connector ``in'' or ``to''. For this reason,  $\lambda$-calculus has also been applied to dependency parse trees in the realm of question-answering systems~\cite{reddy2016transforming}.

\subsection{Types}\label{sec:types}

We now describe a type system that further clarifies the role each entity plays in a hyperedge. In all, we distinguish 8 types, the smallest set we could find that appears to cover virtually all possible information representation roles cataloged in the Universal Dependencies. We first present the types that atoms may have and discuss their use in constructing higher-order entities. We then show how hyperedge types are recursively inferable from the types of the connector and subsequent arguments.

\setlength{\tabcolsep}{4.5pt}

\newcommand{\rcg}{\rowcolor[gray]{.95}}
\renewcommand{\checkmark}{$\times$}
\begin{table*}
\centering
\begin{tabularx}{\linewidth}{ >{\sf}c l X >{\sf\small}l c c }
 \toprule
 \bf{Code} & \textbf{Type} & \textbf{Purpose} & \bf{Example}  & \textbf{Atom} & \textbf{Non-atom}\\
 \midrule
 C & concept & Define atomic concepts & \x{apple/C} & \checkmark & \checkmark \medskip\\
 \rcg{} P & predicate & Build relations  & (\x{is/P} berlin/C nice/C) & \checkmark & \checkmark \\
 \rcg{}M & modifier  & Modify a concept, predicate, modifier, trigger & (\x{red/M} shoes/C) & \checkmark & \checkmark \\
 \rcg{}B & builder   & Build concepts from concepts & (\x{of/B} capital/C germany/C)\quad\quad& \checkmark & \\
 \rcg{}T & trigger & Build specifications & (\x{in/T} 1994/C) & \checkmark & \\
 \rcg{}J & conjunction & Define sequences of concepts or relations & (\x{and/J} meat/C potatoes/C) & \checkmark & \medskip\\
 R & relation & Express facts, statements, questions, orders,...& \x{(is/P berlin/C nice/C)} & & \checkmark \medskip\\
 S & specifier & Relation specification (e.g. condition, time,...) & \x{(in/T 1976/C)} & & \checkmark \\
 \bottomrule
\end{tabularx}
\caption{Hyperedge types with use purposes and examples. Connector types are emphasized with a gray background. The rightmost columns specify whether this type may be encountered in atomic or non-atomic hyperedges.}
\label{tab:types}
\end{table*}

\paragraph{Atomic concepts.}
The first, simplest and most fundamental role that atoms can play is that of a \emph{concept}.  This corresponds to concepts that can be expressed as a single word in the target language, for example ``apple''; they are labeled by this human-readable string, as could be guessed from the previous subsection. 

This defines an eponymous type, ``concept''.
The nomenclature we propose further indicates the type of an atom by appending a more machine-oriented code after this label and a slash (\T{/}). For concepts, this code is ``\T{C}'':
\Q{(apple/C)}

As we shall see, these machine-oriented codes remove ambiguity, facilitate automatic inference and computations. The full list of types as well as their codes and purposes can be seen in table~\ref{tab:types}.

\paragraph{Connectors}
The second and last role that atoms can play is the role of connector. We then have five types of connectors, each one with a specific function that relates to the construction of specific types of hyperedges. 

The most straightforward connector is the {\em predicate}, whose code is ``\T{P}''. It is used to define relations, which are frequently statements. 
Let us revisit a previous example with types:

\Q{(\x{is/P} berlin/C nice/C)}
The predicate \h{is/P} both establishes that this hyperedge is a relation between the entities following it, and gives meaning to the relation. This is isomorphic to typical knowledge graphs~\cite{allen1982s,sowa2014principles} where \h{berlin} and \h{nice} would be connected by an edge labeled with \h{is}.

\medskip 

The \emph{modifier} type (``\T{M}'') applies to one (and only one) existing hyperedge and defines a new hyperedge of the same type. In practice, as the name indicates, it \emph{modifies things} and can be applied to concepts, predicates or other modifiers, and also to triggers, a type that we will subsequently address. For concepts, a typical case is adjectivation, e.g.:

\Q{(\x{nice/M} shoes/C)}
Note here that ``nice'' is being considered as a modifier, while ``nice'' was a concept in the previous case: this is due to the fact that \h{nice/M} and \h{nice/C} refer to two distinct atoms which share the same human-readable label, ``nice''.  To illustrate modification of predicates, let us revisit a previous example, but suppose that we declare that Berlin is not nice. Then we can apply a modifier to the predicate, such as \h{not/M}, so that:
\QC{((\x{not/M} is/P) berlin/C nice/C)}
Finally, modifiers may modify other modifiers:
\QC{((\x{very/M} nice/M) shoes/C)}
The \emph{builder} type (``\T{B}'') combines several concepts to create a new one. For example, atomic concepts \h{capital/C} and \h{germany/C} can be combined with the builder atom \h{of/B} to produce the concept of ``capital of Germany'':
\QC{(\x{of/B} capital/C germany/C)}
A very common structure in English and many other languages is that of the compound noun e.g., ``guitar player'' or ``Barack Obama''. To represent these cases, we introduce a special builder atom that we call \T{(+/B)}. Unlike what we have seen so far, this is an atom that does not correspond to any word, but indicates that a concept is formed by the compound of its arguments; it is necessary to render such compound structures. The previous examples can be represented respectively as \T{(+/B guitar/C player/C)} and \T{(+/B barack/C obama/C)}.

\medskip 
{\em Conjunctions} (``\T{J}''), like the English grammatical construct of the same name, join or coordinate concepts or relations:
\Q{(\x{and/J} meat/C potatoes/C)\\
(\x{but/J} (likes/P mary/C meat/C) (hates/P potatoes/C))}

\noindent We also introduce a special conjunction symbol, \T{(:/J)}, to denote implicit sequences of related concepts. For example, the phrase: ``Freud, the famous psychiatrist'', would be represented as:
\Q{(:/J freud/C (the/M (famous/M psychiatrist/C)))}

\medskip 
The remaining case, {\em triggers} (\T{T}), concerns additional specifications of a relationship, for example conditional (``We go \x{if} it rains.''), or temporal (``John and Mary traveled to the North Pole \x{in} 2015''), local (``Pablo opened a bar \x{in} Spain''), etc.:
\Q{(opened/P pablo/C (a/M bar/C) (\x{in/T} spain/C))}

\paragraph{Hyperedge type inference.}
Atomic types are entirely covered by these six types, of which three exclusively concern atoms (builders, triggers and conjunctions). We already hinted at the fact that non-atomic hyperedges also have types. These are implicit and inferable from the types of the connector and its arguments. Given, for example, that \h{germany/C} is an atom of type concept (\T{C}), the hyperedge \h{of/B capital/C germany/C} is also a concept, and this can be inferred from the fact that its connector is of type builder (\T{B}). Builders need to be followed by at least two concepts. Modifiers (\T{M}) only accept one argument, and the hyperedge in which they participate has the type of the single argument of the modifier. For example, the hyperedge \h{northern/M germany/C} is a concept (\T{C}), and \h{not/M is/P} is a predicate (\T{P}).
 
Table~\ref{tab:type-inference} lists all type inference rules and their respective requirements. They also induce syntactic constraints which close the SH type system.

We may now introduce the two last types of our type system, relation (\T{R}) and specifier (\T{S}), which only concern non-atomic hyperedges: they are always defined as the result of a composition of hyperedges.
\emph{Relations} are typically used to state some fact (even though they can also be used to represent questions, orders and other things). \h{is/P Berlin/C nice/C} is an obvious example of relation. In our context, they thus turn out to be a crucial hyperedge type.  {\em Specifiers} are types that play a more peripheral role, in the proper sense, in that they are supplemental to relations. Specifiers are produced by triggers. For example, the trigger ``\T{(in/T)}'' can be used to construct the specification: \h{in/T 1976/C}. Specifications, as the name implies, add precisions to relations \hbox{e.g.,} when, where, why or in which case something happened.


\begin{table}
\centering\small
\begin{tabular}{ >{\sf}l lc >{\sf}c }
\toprule
 \normalsize\bf{Element types}    &\multicolumn{1}{c}{\normalsize$\rightarrow$} & \normalsize\textbf{Resulting type} \\
 \midrule
 (M \; x) &
 & x\\
 (B \; C \; C+)  && C  \\
 (T \; [CR])  &&  S \\
 (P \; [CRS]+)  && R \\
 (J \; x\; y'+) 
 &&  x \\
 \bottomrule
\end{tabular}
\caption{Type inference rules. We adopt the notation of regular expressions: the symbol $+$ is used to denote one or more 
entities with the type that precedes it, while square brackets indicate several possibilities (for instance, \T{[CR]+} means ``at least one of any of both \T{C} or \T{R}'' types). \T{x} means any type: \T{(M x)} is of type \T{x}.}
\label{tab:type-inference}
\end{table}

\subsection{Argument roles}\label{sec:argroles}

We introduce a last notion that we employ to make meaning more explicit: \emph{argument roles} for builders and predicates. They are represented as sequences of characters that indicate the role of the respective arguments following such connectors.

\paragraph{Concept builders.}
Given a concept hyperedge, a key issue is that of inferring its \emph{main concept}, \hbox{i.e.} the concept that can be assumed to be its hypernym.  Beyond the simple case of atoms, concept hyperedges may only be formed by connectors that are either modifiers or builders. When the connector is a modifier, finding the hypernym is admittedly trivial.
When the connector is a builder, it is often possible to infer the main concept among the arguments. There are only two possible roles: ``main'' (denoted by \T{m}) and ``auxiliary'' (denoted by \T{a}).  For example: 
\QC{(+/B.am tennis/C ball/C)}
The argument role annotation ``\T{.am}'' indicates that \T{ball/c} is the main concept in the construct, meaning that \T{(+/B.am tennis/C ball/C)} is a sort of \T{ball/c} --- the main concept is a hypernym of the whole construct.

With compound nouns (\h{+/B} builder), we simply make use of part-of-speech and dependency labels to infer the main concept.  Another common situation where finding roles is quite trivial is the case of builders derived from a proposition, such as \h{of/B}, which express a relationship between the arguments. For example, in \h{of/B.ma capital/C germany/C}, the main concept is \h{capital/C}. ``Capital of Germany'' is thus a type of capital. In English and many other languages, it is always the case that the main concept is the first argument after a builder derived from a proposition.

\paragraph{Predicates.}

Predicates can induce specific roles that the following arguments play in a relation.
The need for argument roles in relations arises from cases where the role cannot be inferred from the type of the argument. For example, the same concept could participate in a relation as a subject or as an object.
Consider for instance the sentence ``John gave Mary a flower'', represented as:
\Q{(gave/P.sio john/C mary/C (a/M flower/C))}
In this relation, the argument role string ``\T{sio}'' indicates that the three arguments following the predicate respectively play the roles of subject, indirect object and direct object. 
This relation involves three concepts united by the predicate that represents the act of giving, but without the argument roles, who the giver is, who the receiver is, and what object is being given, would remain undefined. Relying on ordering would not be enough, both due to the flexibility of NL in this regard, and to the fact that the presence of a certain role after a predicate is often optional. 

There are admittedly more possible roles than for builders. They are shown in table~\ref{tab:pred-argument-roles}. Once again, this set is the result of an effort to cover all  grammatical cases listed in the Universal Dependencies in the most succinct way possible. Most of them (in fact, the first $8$ in the table) directly correspond to generic grammatical roles of the same name. Of these, the first $6$ are by far the most frequent.
\emph{Specifications} were already discussed in the previous subsection (\ref{sec:types}),
and their purpose as hyperedges coincides with their role when participating in relations: as an additional specification to the relation (temporal, conditional, etc.).
Finally, a \emph{relative relation} is a nested relation, that acts as a building block of the outer relation that contains it. We will make extensive use of this later, to identify what is being claimed by a given actor.

\begin{table}
\centering
{
\begin{tabular}{ l >{\sf}c }
 \toprule
 \textbf{Role} & \textbf{Code} \\
 \midrule
 active subject  &  s \\
 passive subject  & p \\
 agent (passive)  & a \\
 subject complement & c \\
 direct object  &  o \\
 indirect object  &  i \\
 parataxis & t \\
 interjection & j \\
 specification & x \\
 relative relation & r \\
 \bottomrule
\end{tabular}
\caption{Predicate argument roles.}
\label{tab:pred-argument-roles}
}
\end{table}

\section{Translating NL into SH}
\label{sec:text2hyper}

We now discuss the crucial task of translating NL into this SH representation. This can, of course, be framed as a conventional supervised ML task. A difficulty arises from the lack of training data. SH is a novel representation, and the effort necessary to annotate a sufficiently large amount of text to train an NL to SH translator from scratch is far from trivial. We were motivated to look for an alternative, and we hypothesized that it would be much easier to infer the SH representation from grammatically-enriched representations than from raw text. We will show that this indeed appears to be the case.

We propose a two-staged approach. The first ($\alpha$-stage) is a classifier that assigns a type to each token in a given sentence. The second ($\beta$-stage) is a search tree-based algorithm that recursively applies the rules in table~\ref{tab:type-inference} to impose the hypergraphic structure on the sequence of atoms produced by the $\alpha$-stage. This restricts the ML part of the process to the $\alpha$-stage, making it a trivial classification problem. 

\subsection{$\alpha$-stage}
The classification categories correspond to the set of the six atomic types shown in table~\ref{tab:types}, with one additional category for tokens that should be discarded (typically punctuation). The open question is the feature set. We will see how, operating on the previous assumption regarding grammatical annotation, we use spaCy\footnote{ An open-source library for NLP in Python which includes convolutional neural network models for tagging, parsing and named entity recognition in multiple languages. A relatively recent comparison of ten popular syntactic parsers found spaCy to be the fastest, with an accuracy within 1\% of the best one~\cite{choi2015depends}} -- a popular NLP tool -- to generate appropriate features.

Using this library we perform segmentation of text into sentences, followed by tokenization and annotation of tokens with parts-of-speech, dependency labels and named entity categories. In short, we deploy the full arsenal of off-the-shelf NLP tasks that come available with spaCy. In this work we restrict ourselves to the English language and we use the ``en\_core\_web\_lg-2.0.0'' language model.

We collected randomly selected texts in English from five categories: fiction (5 books, 87738 sentences) and non-fiction books (5 books, 51597 sentences), news (10 articles, 532 sentences), scientific articles (10 articles, 3467 sentences) and Wikipedia articles (10 articles, 2888 sentences). From these we selected 60 random sentences in each category, thus a total of 300 sentences representing 6936 tokens. An interactive computer script was used to aid in the process of manually annotating each word 
of these sentences with one of the \emph{alpha} categories i.e., atomic types. 
These were used to train a random forest classifier. For this purpose we employed the one included with \emph{scikit-learn} (version 0.23.2), a widely used ML package. We did not perform any hyperparameter tuning, and used the default parameters set by this version of the package. There is possibly room for improvement here. For the aims of this work, we found it preferable to avoid introducing potentially confounding factors that could arise from hyperparameter optimization efforts.

\paragraph{Feature definition.} We consider an initial set that encompasses all the potentially useful features that we could derive from a standard NLP pipeline such as spaCy. As we mentioned, it provides dependency parse labels (referred to, from now on, as \emph{DEP}) and named entity recognition categories (\emph{NER}). Parts-of-speech are provided in two flavors: the more extensive OntoNotes tag set (version 5) from the Penn Treebank, and the simpler Universal Dependencies (UD) part-of-speech tag set (version 2). 
Accuracy values for each of these elements are reported in \cite{spaccuracy} to be $0.97$ for the fine grained part-of-speech tagger (i.e., guessing the OntoNotes tag), $0.92$ for unlabeled dependencies (i.e., guessing the head of each token) and $0.90$ for labeled dependencies (i.e., the head and the label).

Let us refer to the former as \emph{TAG}, and to the latter as \emph{POS}. We can also consider the most common words in the corpus. We consider as features the sets of 15, 25, 50 and 100 most common words (\emph{WORD15}, \emph{WORD25} and so on). Further features indicate if a token corresponds to some punctuation symbol, if it is at the root of the dependency parse trees, if it has left or right children in this same tree, and finally its shape in terms of capitalization (e.g. the shape of the word ``Alice'' is Xxxxx). Then, we establish three types of relative tokens: the ones that appear directly after or before the current one in the sentence, if they exist, and the one that is the parent of the current one in the dependency parse tree, if it exists. For each one of these tokens, all the previous features are also applied (for example, the UD part-of-speech of the dependency head is \emph{HPOS}, and the part-of-speech of the subsequent word in the sentence is \emph{POS\_AFTER}). We thus have $33$ candidate features in total. All of these features are categorical, and we employ one-hot encoding to feed them to the decision trees.

\paragraph{Feature selection.}
We tested two approaches for feature selection: a very simple genetic algorithm (GA) and iterative ablation. For the GA, we encoded features as bits (acting as switches to specify which features belong to the set). We used mutation only (bit-flip with a probability of $.05$), a population of $100$, and parent selection through a tournament of $3$. Search stopped at $100$ generations without improvement. The fitness function was the mean of $5$ evaluations of the accuracy of the feature set, each with a distinct and randomly selected split of the training / testing data. This eventually resulted in a set of $15$ features: \{\emph{WORD25}, \emph{TAG}, \emph{DEP}, \emph{HWORD25}, \emph{HWORD50}, \emph{HWORD100}, \emph{HPOS}, \emph{HDEP}, \emph{IS\_ROOT}, \emph{NER}, \emph{WORD\_BEFORE15}, \emph{WORD\_BEFORE100}, \emph{WORD\_AFTER15}, \emph{PUNCT\_BEFORE}, \emph{POS\_AFTER}\}.

The iterative ablation procedure starts with the set of all candidate features, and $100$ runs of the learning algorithm are performed, again each run randomly split into two-thirds for training and one-third for testing. This provides us with a set of $100$ accuracy measurements. The process is then repeated, excluding one feature at at time. The feature that most degrades mean accuracy is excluded. If no feature has a negative impact on accuracy, then the one with the highest p-value (according to the non-parametric Kolmogorov–Smirnov test) above a threshold is excluded. The procedure is repeated, ablating one feature at a time, until no remaining feature fulfills any of the previous two criteria. We performed this procedure with threshold p-values of $.05$ and $.005$. The first left us with a set of five features: F5 = \{\emph{TAG}, \emph{DEP}, \emph{HDEP}, \emph{HPOS}, \emph{POS\_AFTER}\}; the second with three features: F3 = \{\emph{TAG}, \emph{DEP}, \emph{HDEP}\}.

\begin{figure*}[!t]
\centering
\includegraphics[width=\linewidth]{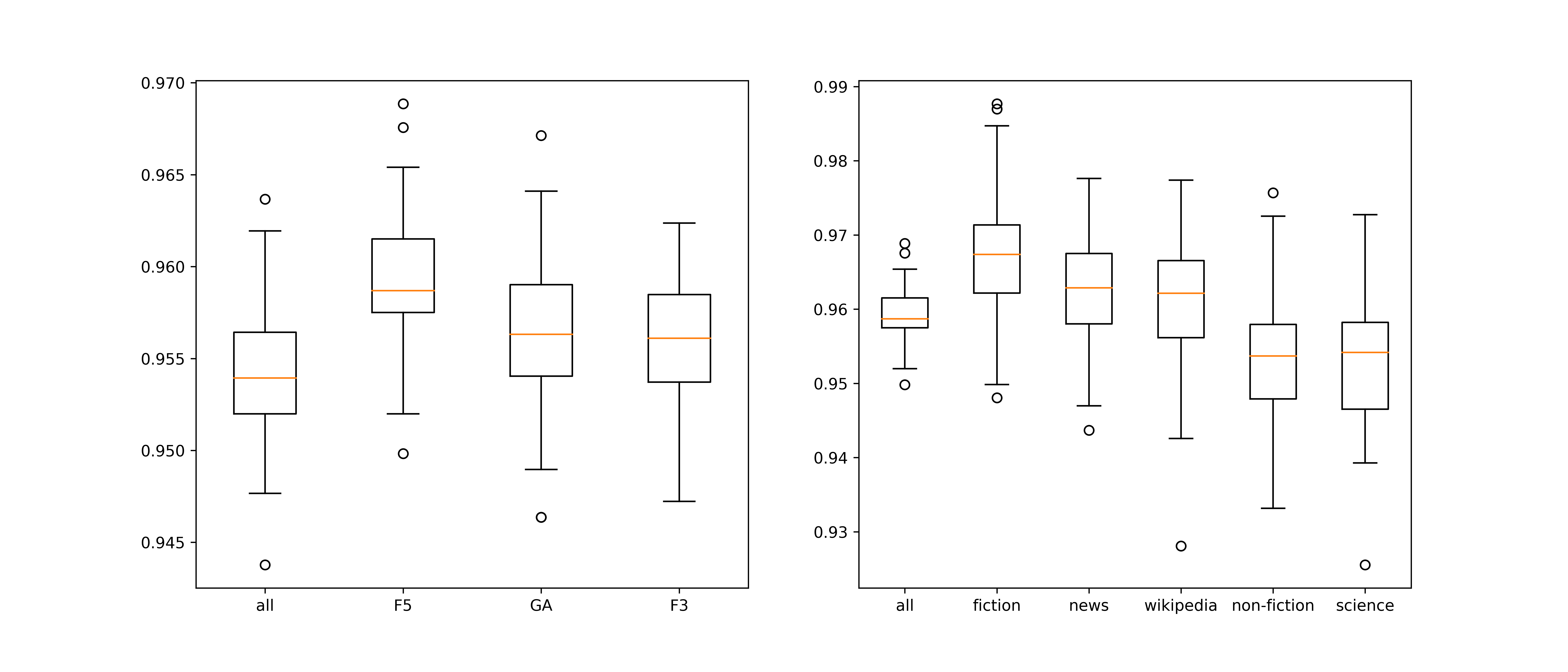}
\vspace{-2em}
\caption{\emph{Left:} accuracy of the $\alpha$-classifier, comparing several feature sets; \emph{all} includes all features, \emph{GA} a features set obtained with a genetic algorithm, {F3} is the outcome of iterative ablation with $p < .005$ and {F5} with $p < .05$. \emph{Right:} accuracy by source text category using {F5}.}
\label{fig:alpha-features-categories}
\end{figure*}

The results of these experiments are shown on the left side of figure~\ref{fig:alpha-features-categories}. As can be seen, all of the three attempts outperform the set of all features. Interestingly, {F5} is significantly better than {F3}, even at $p < .005$. The accuracy of the GA set falls between that of {F3} and {F5}. We performed these experiments not only as an endeavor to achieve acceptable accuracy for the experiments that follow, but also to obtain empirical evidence regarding the relationship between SH types and traditional linguistic features. We can conclude that SH does not correspond to some trivial mapping of any single linguistic feature. For subsequent experiments we will use {F5}, given that it has the best accuracy and still uses a relatively small number of features -- something that can make a difference regarding the computational effort needed to parse large quantities of text. It is interesting to notice that {F3} still leads to a higher accuracy than the set of all features, and having only three features, such a classifier could be feasibly implemented in a purely programmatic way. A completely human-understandable classification tree could be produced, and also implemented in a very efficient way, sacrificing relatively little in terms of accuracy.

On the right side of figure~\ref{fig:alpha-features-categories} we present the accuracy of the classifier by text category, using {F5}. Here, it is interesting to note that the best performing category (fiction) and also one of the second-best (wikipedia, which is not significantly different from news) are out-of-corpus for the training set of the ML model of the underlying linguistic features.  It is remarkable that the accuracies that we achieve are comparable and may even surpass the values reported by spaCy (see above). In other words, this suggests that, far from accumulating errors down the stream of the various processing steps, our $\alpha$ stage appears to even correct upstream errors.

It is conceivable that more features become relevant, if a larger number of exotic cases becomes available through larger training corpora. It is also conceivable that larger windows (beyond just previous and next token) become relevant with larger datasets and more sophisticated ML approaches. Such considerations are beyond the scope of this work.

\subsection{$\beta$-stage}

The $\beta$-stage transforms the sequence of atoms of the original sentence, each typed by the $\alpha$-stage, into a semantic hyperedge that reflects the meaning of the sentence and respects the SH syntactic rules. In practice, this operation amounts to a bottom-up process that aggregates the deeper structures of the sentence into increasingly complex hyperedges, by recursively combining them until only a final, well-formed semantic hyperedge is left.

\SetKwProg{Fn}{Function}{}{end}\SetKwFunction{FBeta}{BetaTransformation}
\SetKwProg{Fn}{Function}{}{end}\SetKwFunction{FApply}{ApplyPattern}%
\SetAlgoLongEnd

\newcommand{\forcond}{$i=0$ \KwTo $n$}

\newcommand\doubleplus{+\kern-1.3ex+\kern0.8ex}

\begin{algorithm}[!th]\small
\DontPrintSemicolon

\Fn(){\FApply{seq, pos, pat}}{
    \KwData{A sequence of edges $seq$, a position in the sequence $pos$ and a pattern $pat$}
    \KwResult{A sequence of edges with the initial edges replaced by a single one, if they match the pattern.}

    \uIf{$pat$ matches $seq$ at $pos$}{
        $edge \longleftarrow$ reorder matching elements of $seq$ to align with $pat$ \;
        $seq' \longleftarrow$ matching part of $seq$ replaced with $edge$ \; 
        \KwRet $seq'$
    }
    \Else{
        \KwRet $\varnothing$
    }
}
\;
\Fn(){\FBeta{seq}}{
  \KwData{A sequence of edges $seq$}
  \KwResult{An edge $e$}
  
  \If{$|seq| = 1$}{
    \KwRet $seq[0]$ \;
  }
  
  $heu_{best}  \longleftarrow -\infty$ \;
  $seq_{best}  \longleftarrow \varnothing$ \;
  \For{$pos=1$ \KwTo $|seq|$ }{
    \For{$pat \in Patterns$}{
        $seq' \longleftarrow$ \FApply{seq, pos, pat} \;
        $heu  \longleftarrow h(seq, pos, pat)$ \;
        \If{$seq' \ne \varnothing \land heu > heu_{best}$}{
            $heu_{best}  \longleftarrow heu$ \;
            $seq_{best}  \longleftarrow seq'$ \;
        }
    }
  }
  \uIf{$seq_{best} \ne \varnothing$}{
    \KwRet \FBeta{$seq_{best}$} \;
  }
  \Else{
    \KwRet (\h{:/J} $\doubleplus$ $seq[:2]$ ) $\doubleplus$ $seq[2:]$\;
  }
}
\label{algo:beta_stage}
\caption{The $\beta$ transformation recursively applies the patterns from type inference rules until only the final hyperedge is left.}
\end{algorithm}

The process for this transformation is formalized in algorithm~\ref{algo:beta_stage}. Let us nonetheless explain in plain words how $\beta$ iteratively constructs a hyperedge, which need not be a proper semantic hyperedge except at the final step. The process starts indeed with an initial hyperedge as the simple sequence of typed atoms of the original sentence. At each step, the elements of the currently-formed hyperedge are scanned from left to right to look for a sub-sequence of types that matches the list on the left side of the type inference rules of table~\ref{tab:type-inference}, taken as unordered patterns i.e., up to any reordering. For instance, ``capital of Germany'' may have been parsed by $\alpha$ as a typed sub-sequence ``\T{capital/C, of/B, germany/C}'', which then matches the second pattern \T{(B C C)}. It may then be rearranged as such by putting the connector in first position and preserving the order of the remainder of the hyperedge i.e., ``\T{(of/B capital/C germany/C)}'', which conforms to the second inference rule of table~\ref{tab:type-inference}.  
Note that, in practice, we also restrict the second and fifth patterns, i.e. the builder and conjunction patterns, to the minimum number of two arguments: respectively \T{(B C C)} and \T{(J $x$ $x'$)}. We find that it fits NL more naturally and thus leads to more correct parses. Further tasks of knowledge inference might later introduce builder- and conjunction-based structures with more arguments. 
We complement the patterns with one rule that corresponds to the special connector \T{(+/B)}. This extra rule is admittedly needed to transform implicit builders \T{(C C)} into \T{(+/B C C)}.

If only one sub-sequence matches, it is transformed into a sub-hyperedge by application of the rule. 
If two or more sub-sequences match, the $\beta$-stage needs to make a decision on which one to choose and proceed with as if only one sub-sequence matched. For this case, we use a heuristic function (this is function $h$ in algorithm~\ref{algo:beta_stage}). This heuristic function relies on the grammatical structure of the sentence given by the dependency tree. Our hypothesis is that grammatically connected edges are more likely to belong to the same higher-order edge, so the first criterion of $h$ is to always assign a higher score to sub-sequences where all items are directly connected in the dependency tree. By ``directly connected in the dependency tree'', we mean that all hyperedges contain one atom/token that is the head or the child of at least one atom/token in another hyperedge, and that any hyperedge can be reached from any other, following such grammatical links. In case there is a tie, the heuristic function then prefers the sub-sequence that contains the deepest atom/token in the dependency tree -- again assuming a correlation with SH depth, and thus respecting the bottom-up process of the $\beta$-transformation. Finally, if there is still a tie, rules are applied by the order of priority expressed in table \ref{tab:type-inference}, which is empirically organized by decreasing order of the depth at which each respective structure tends to appear in hyperedges. The special rule for \T{(+/B)} is assigned the highest priority.

If no sub-sequence matches, the two first items in the sequence are connected by prepending the special conjunction \T{(:/J)}, which is meant to convey the most generic and abstract meaning of ``these two things are related in the most generic sense''. This captures cases often found in natural language, such as: ``A new era: quantum computation is here.'', which translates to:
\Q{(:/J (a/M (new/M era/C)) (is/P (quantum/M computation/C) here/C))}

If the resulting hyperedge entirely conforms to one of the type inference rules, the process stops successfully as it managed to form a recursively correct semantic hyperedge. Otherwise, the process is reiterated on the newly-formed hyperedge. 
The process is thus guaranteed to converge on a syntactically valid hyperedge, but is of course not guaranteed to produce the most desirable or correct representation.
However, we experimentally verify below that, given a correct classification from the $\alpha$-stage and a correct dependency parse tree, this process consistently leads to the construction of a SH that correctly conveys the meaning of the original sentence.

\begin{figure*}[!t]
\centering
\includegraphics[width=\linewidth]{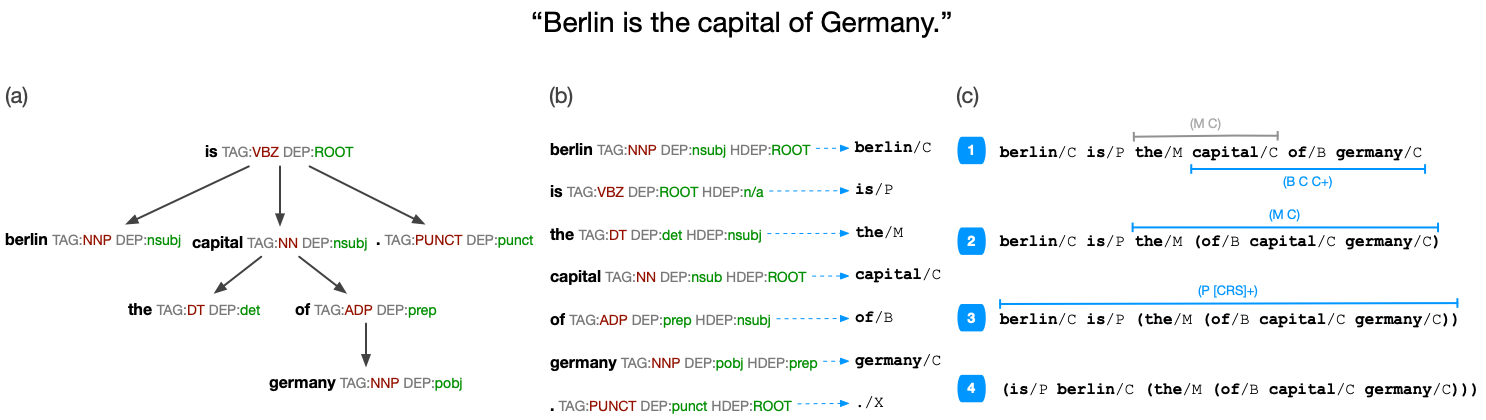}
\caption{{\textbf{(a)}} Dependency parse tree with dependency labels (green) and fine grained part-of-speech tags (red). \textbf{(b)} $\alpha$-stage classification of atom types. \textbf{(c)} $\beta$-stage structuring of sentence by iterative application of the patterns from table~\ref{tab:type-inference}. A non-selected pattern is greyed-out.}
\label{fig:NL2SH}
\end{figure*}

Let us first illustrate the $\beta$-stage in figure~\ref{fig:NL2SH}, which provides one example of an entire parsing process (using the \emph{F3} feature set for simplicity). In figure~\ref{fig:NL2SH}(c), the recursive application of $\beta$-transformations to an initial sequence of atoms can be followed. In the first step, we can see that the sequence \T{(the/M, capital/C)} matches the pattern \T{(M C)}, and the sequence \T{(capital/C, of/B, germany/C)} matches the pattern \T{(B C C+)}. We thus rely on the above-mentioned heuristic function, which causes \T{(of/B capital/C germany/C)} to be preferred to \T{(the/M capital/C)}. The reader can verify that selecting the latter at this stage would lead to a dead-end. The rest of the SH construction is straightforward.

\begin{table*}[!t]
\centering
\begin{tabular}{ l | c c c c | c }
\toprule
 \textbf{Category} & \textbf{Correct} & \textbf{Defect}
 & \textbf{Wrong} & \textbf{Total} & \textbf{Mean relative defect size} \\
 \midrule
Non-fiction & 87 (.87) & 8 (.08) & 5 (.05) & 100 & .188 \\
Wikipedia & 81 (.81) & 12 (.12) & 7 (.07) & 100 & .190 \\
News & 77 (.77) & 16 (.16) & 7 (.07) & 100 & .147 \\
Fiction & 79 (.79) & 5 (.05) & 16 (.16) & 100 & .140 \\
Science & 71 (.71) & 19 (.19) & 10 (.10) & 100 & .290 \\
\textbf{All} & \textbf{395 (.79)} & \textbf{60 (.12)} & \textbf{45 (.09)} & \textbf{500} & \textbf{.206} \\
 \bottomrule
\end{tabular}
\caption{Global NL to SH parser evaluation.}
\label{tab:parser-evaluation}
\end{table*}

\paragraph{Argument roles.} Now that the core of the translation of NL into SH has been specified, assigning the argument roles introduced in Section~\ref{sec:argroles} amounts to a trivial translation from the dependency labels. Sometimes however, the parser may fail to determine an argument's role, and thus classify it as \emph{unknown} (that we code ``\T{?}'' for this purpose).

\subsection{Validation of $\alpha$ and $\beta$}

To test the accuracy of the complete translation from NL to SH, we randomly selected $100$ new sentences for each text category, that were used neither for training nor testing of the $\alpha$-classifier. We establish three categories: completely correct hyperedges, hyperedges with some defect and completely wrong hyperedges. A hyperedge is considered to have a defect if overall meaning is preserved, but some subedge contains a defect. Let us consider a real example from our dataset. The sentence: ``The scientists – who are part of a multi-year International Shelf Study Expedition – stressed their findings are preliminary.'' was parsed as:
\Q{(stressed/P (:/J (the/M scientists/C) (are/P who/C (of/B part/C (a/M (+/B (+/B (+/B multi/C -/C) year/C) (+/B international/C (+/B shelf/C (+/B study/C expedition/C)))))))) (are/P (their/M findings/C) preliminary/C))}

The hyperedge preserves most of the meaning of the sentence, but the concept \T{(+/B (+/B multi/C -/C) year/C)} is not correctly formed. Either \T{(-/B multi/C year/C)} or \T{(multi/M year/C)} would be much preferable. However, this partially defective parse is still likely to be useful in the methods that we will discuss in the following sections. We also see how different type assignments of the $\alpha$-classifier can result, in practice, in correct hyperedges at the end. We can also use this example to illustrate another metric that we employ in this evaluation: the relative defect size. This is simple the ratio of the size of the defective part to the size of the entire hyperedge. Size is measured in total number of atoms (at any depth).

A wrong hyperedge is one where the meaning of the sentence is completely lost. For example, consider what would happen if, in the above case, ``stressed'' was classified as a concept instead of a predicate. This also serves to illustrate that there is a complex relationship between $\alpha$-classifier accuracy and overall parser accuracy. Some mis-classifications at the $\alpha$-stage can still allow for a completely correct parse, while others can lead to catastrophic failures or just minor defects. Nonetheless, we observe on this sample of 500 sentences that a correct $\alpha$ classification and dependency parse tree always lead to the construction of an SH that preserves the meaning of the sentence. By contrast, a badly-structured dependency tree appears to have a significant negative impact on the functioning of $\beta$, through the heuristic function. If this result generalizes, this suggests that, for a given accuracy of the dependency parsing module, increasing the quality of the NL to SH translation principally relies on improving $\alpha$ and the heuristic function.

We show the results of this evaluation in table~\ref{tab:parser-evaluation}. It is interesting to notice that ``non-fiction'' is one of the worst performing categories in the $\alpha$-classifier, but ends up being the best one overall. Likewise, ``fiction'' is the best category at $\alpha$-stage but ends up being the second worst here. Unsurprisingly, ``fiction'' sentences tend to be richer in figures of speech and other complexities and ambiguities that lead to a higher rate of catastrophic failure. Conversely, ``non-fiction'' is the category with the most straight-forward sentences. In the ``science'' category, the difficulties are more related to a variety of unusual technical terms and notations, that lead more to defects than catastrophic failures. Overall, we see that a high percentage of the texts are correctly translated to SH, even in the worst-performing categories.

We only work with English in this article, but supporting a new language essentially requires to generate a relatively small number of $\alpha$-classifier training examples. The rest of the process is currently language-agnostic: even though more research would be needed to explore this issue thoroughly, the fact that we cover all of the Universal Dependency cases gives us good reasons to believe that NL to SH translation shall be applicable to any language. The software package that we released to implement all the ideas discussed in this article includes the interactive script that we used to perform this annotation task ourselves.

\section{Knowledge Inference and Extraction}
\label{sec:knowledge-inference}

We are finally in the position to explore the use of SH extracted from open text to perform language understanding tasks. First we will discuss how we naturally generalize SH to represent patterns and inference rules, and then we will manually define three such rules to perform conjunction resolution: a very useful and generic task that will be used in every following practical application discussed in this work, and very likely useful for myriad knowledge inference and extraction tasks. Then we will discuss how we systematized the process of discovery of useful patterns, and how we use this process to discover $8$ patterns for the purpose of Open Information Extraction (OIE), for which an abundant computer science literature exists where scholars are interested in inferring relations from free text. We will demonstrate the expressive power of SH by showing that these simple patterns produce competitive results when compared with a number of contemporary systems targeted at OIE, using an external benchmark.

\subsection{A pattern-matching language}

From a text corpus, the NL to SH translation stage attempts to convert each sentence into a hyperedge. In practice, all resulting hyperedges are stored in a proper SH database. From there, language understanding tasks may be performed in the form of inferences, which we define using SH notation with the help of patterns. Broadly, inference rules and patterns may also be written as hyperedges.

\paragraph{Variables and patterns.}
We introduce the concept of \emph{variable}. A variable simply indicates a placeholder that can match a hyperedge, and then be used to refer to that hyperedge. Unlike the other atoms we have seen so far, variables are represented in capital letters. With variables we can define \emph{patterns}, that can then be matched against other hyperedges. For example, consider the pattern:
\QC{(is/P.sc SUBJ PROP/C)}
\noindent which matches, for example:
\QC{(is/P.sc (the/M sky/C) blue/C)}

\newcommand{\imply}{\:\mathbf{\vdash}\:}
Notice that the variable PROP includes a type code, while the predicate ``\T{is/P.sc}'' features argument roles. If type codes or argument roles are added to variables, this simply means that a hyperedge only matches this variable if the types and argument roles also match. 

We introduce a few more notation details for argument roles in patterns. In practice, we often allow the various pattern elements to appear in any order, denoting the order-indifferent roles between curly brackets ``\{ \: \}''. The arguments can appear in any order, as long as all of the pattern roles are present. For instance,
\QC{(is/P.\{sc\} SUBJ PROP/C)}
would both cover \T{(is/P.sc SUBJ PROP/C)} and \T{(is/P.cs PROP/C SUBJ)}.
Furthermore, it is possible to specify the optional presence of certain arguments by listing them as ``...'', which simply indicates that any number (including zero) hyperedges may be present at that point.  In a pattern, if the connector indicates argument roles, then any further arguments may be present, unless indicated otherwise.  In case connectors do not indicate argument roles, ``...'' can thus be used to indicate that more hyperedges at a certain point are permissible. For instance, \QC{(is/P.\{sc\} SUBJ PROP/C ...)}
matches ``The sky is blue today'' and ``Today the sky is blue''.
It is also possible to denote an undefined sequence of hyperedges with a variable name by using ``X...'', which thus refers to the same specific sequence everywhere it is used. 

Sequences of alternative arguments roles of which anyone of them can be matched once are represented inside square brackets. For example, ``\T{[sp]}'' matches either a subject or a passive subject, once. Finally, 
it is possible to forbid the presence of arguments with a certain role by listing them after ``-''. For example, in the pattern ``\T{(PRED/P.-sp X...)}'', arguments with roles ``\T{s}'' or ``\T{p}'' are not allowed; it would however match \T{(play/P.o football/C)}. 



\begin{table*}[!h]
\begin{center}
\begin{tabular}{ c >{\sf}l c }
\toprule
 \textbf{\#} & \bf{Rule} & \textbf{Inferences} \\
 \midrule
1 & $\big($*/J \, ... \, CONCEPT/C \, ...$\big)$ $\imply$ (CONCEPT/C) & 147 \\
\rcg{}2 & $\Big($*/J \, ... \, $\big($PRED/P.\{[sp]\} \, X \, Y...$\big)$ \, ...$\Big)$ $\imply$ $\big($PRED/P.\{[sp]\} \, X \,Y...$\big)$& 63 \\
3 & $\Big($*/J \, $\big($*/P.\{[sp]\}\, SUBJ/*\, ...$\big)$ \, ... \, $\big($PRED/P.-sp \, X...$\big)$ \, ...$\Big)$ $\imply$ $\big($PRED/P.\{s\} \, SUBJ/* \, X...$\big)$ & 10\\
 \bottomrule
  \end{tabular}
\caption{Conjunction resolution rules and respective number of inferred hyperedges from the OIE benchmark}.
\label{tab:conjunction-rules}
\end{center}
\end{table*}

\paragraph{Rules.}
We may now define \emph{rules} which we denote with a couple of patterns separated by the symbol ``$\vdash$'', as in ``\T{PATTERN1~$\vdash$~PATTERN2}''. This notation indicates that any hyperedge that contains a hyperedge matching the left-hand-side \T{PATTERN1} would incur the creation of a duplicated hyperedge consisting of the matching portion rewritten according to the right-hand-side expressed as \T{PATTERN2}. In a sense, these are replacement rules, except that the original hyperedge is preserved.

For example, consider the rule:

\Q{(is/P.sc SUBJ PROP/C) $\imply$ (property/P PROP)}

\noindent which, applied to the above example, produces the inference:

\Q{(property/P blue/C)}

In essence, a rule makes it possible to populate an SH database with new knowledge that is inferred from NL yet need not, in turn, correspond to an actual sentence.


\subsection{Conjunction Decomposition}
\label{sec:conjres}

Decomposing relations that include conjunctions into simpler relations not only facilitates OIE tasks, but is also of general usefulness in knowledge inference tasks. We show the three rules that we developed manually to perform conjunction decomposition in table~\ref{tab:conjunction-rules}.



The first rule concerns conjunctions of concepts, such as ``Mary likes books and flowers.'':
\QC{(likes/P.so mary/C (and/J books/C flowers/C))}
where we generate one relation for each element:

\Q{(likes/P.so mary/C books/C)\\
(likes/P.so mary/C flowers/C)}

The second rule concerns conjunctions of relations with explicit subjects, for example: ``Mary likes astronomy and Alice plays football.'', which is parsed as:
\QC{(and/J (likes/P.so mary/C astronomy/C) (plays/P.so alice/C football/C))}
\noindent is decomposed into: ``Mary likes astronomy.'' and ``Alice plays football.'' i.e.,:
\Q{(likes/P.so mary/C astronomy/C)\\
(plays/P.so alice/C football/C)}

The third rule makes the subject explicit in situations such as ``Mary likes astronomy and plays football.'' i.e.,
\QC{(and/J (likes/P.so mary/C astronomy/C) (plays/P.o football/C))}
\noindent inferring that ``Mary'' is the subject from the first relation in the conjunction and applying it to the second one: ``Mary plays football.'', resulting in:
\Q{(likes/P.so mary/C astronomy/C)\\
(plays/P.so mary/C football/C)}
\noindent In practice, this is done by remembering the last argument with the subject (\emph{s}) role and applying it to the following relations that miss a subject.

\smallskip 
Naturally, these rules have a lot of space for improvement, not making distinctions for conjunctions with special meaning (e.g. ``but'', ``instead'', etc.). Nevertheless, we will see that they are already quite successful in the 
tasks that we will subsequently present.

\subsection{Pattern Learning}
\label{subsec:pattern-learning}

\begin{figure*}[!t]
\centering
\includegraphics[width=\linewidth]{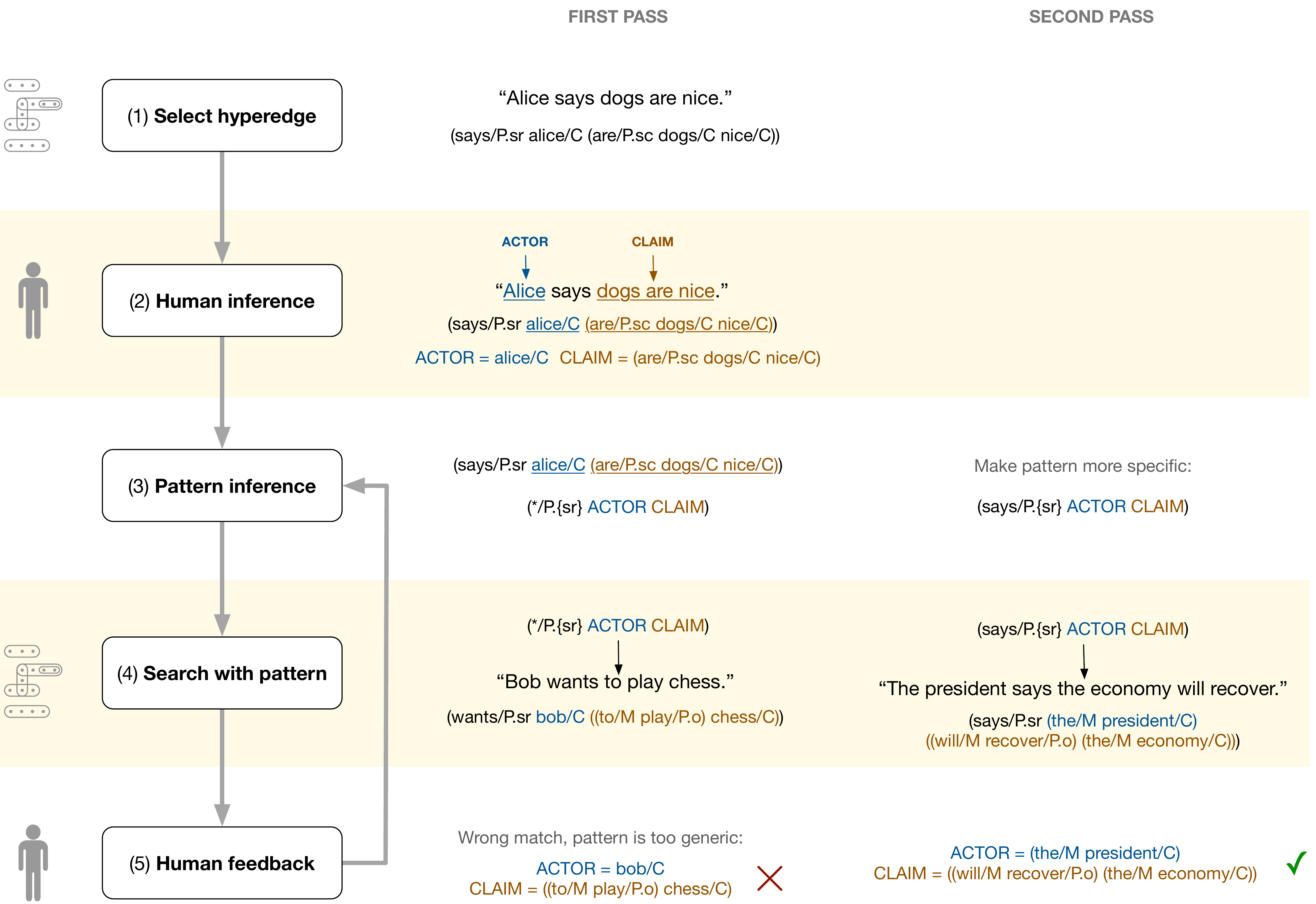}
\caption{Pattern learning template and example with two passes. At the end of the second pass, the pattern \h{says/P.{sr} ACTOR CLAIM} is confirmed to work.}
\label{fig:pattern-learning}
\end{figure*}

With the help of 
hypergraphs extracted from corpora of open text, it becomes possible to define a systematic process of discovery of patterns that enable knowledge extraction, with a human-in-the-loop. On the left side of figure~\ref{fig:pattern-learning} we present the general template for such a process. We will use this template to illustrate both how we discovered the patterns for the OpenIE task as well as the claim and conflict analysis that will be presented in section~\ref{subsec:openie}, and also how more sophisticated and automated pattern learning systems can be created.
The rest of figure~\ref{fig:pattern-learning} is a step-by-step illustration on a simple example aimed at generating patterns to detect claims.

In step (1), a hyperedge is selected from the hypergraph generated from a given training corpus. It can be drawn at random, or by any other criterion adapted to the pattern-learning task at hand. For instance, if we want to learn patterns typical of claims, we can first focus on hyperedges starting with a predicate (\T{*/P}) and, more precisely, the most frequent ones among them, as a strategy to attain good coverage. We observe that ``\T{says/P}'' is such a predicate and based on this, we draw 
\T{(says/P.sr alice/C (are/P.sc dogs/C nice/C))} i.e., ``Alice says dogs are nice''.  Other pattern-learning tasks may naturally require different selection criteria, and in the next subsection~(\ref{subsec:openie}) we will provide another and more general example.


A human is then presented with this hyperedge in step (2), and asked to manually perform an inference. The inference consists of selecting sub-edges and assigning them to variables according to some schema. In the example shown, the inference aim is to detect which actors making claims, and thus identify which ``\T{ACTOR}'' makes which ``\T{CLAIM}''.

Step (3) consists of generalizing the original hyperedge into a pattern with the help of the variable assignments. The idea is to create the most generic pattern that fits the human inference. The matching parts are replaced by the corresponding variables, and the remaining sub-edges are replaced by wildcards, while maintaining type annotations. 

Then the process goes back to the whole training hypergraph. Step (4) consists of finding hyperedges that match the pattern, so that they can be presented to the human for validation. Then, in step (5), the human can simply indicate if these further matches are valid or not.

When a match is not valid, the process can then return to step (3) and use this information to refine the pattern. What is now the most general version of the pattern that does not match the previously detected incorrect case?

In our work, we used a conventional Jupyter notebook directly accessing the programmatic interface of Graphbrain\footnote{https://github.com/graphbrain/graphbrain}, the open-source library that we developed to implement all of the ideas discussed in this work. Refinements at step (3) were performed manually, testing hypothesis on the most general version of a pattern by simply asking Graphbrain to check how many actual hyperedges match each attempt, and choosing the one with the highest value. This process can obviously be automated with a search tree, that attempts a number of substitutions -- more or less generic wildcards, lemma matching, atom root matching, structural matching, etc. -- at each step, and then uses the training hypergraph to empirically test them and discover the most generic one that correctly matches both the positive and negative cases known so far. Then, even more sophisticated possibilities arise, such as the integration with general knowledge databases (e.g. specifying that a variable must be a concept of type ``country'', or that a predicate must be the synonym of a certain action), or the use of auxiliary methods such as semantic proximity with word2vec-like embeddings, or hybridization with ML algorithms.

Another possible improvement is in the domain of software and user-interface development, allowing for less technical users to provide inferences and feedback -- a user can be invited to directly select parts of a sentence and assigning them to meanings (e.g. ``actor'', ``claim'', ``aggressor'', etc.), without having to see or interact with hypergraphic notation. Such refinements are beyond the scope of this work, but it is our hope to lay the foundations for these and other possibilities.

\subsection{Open Information Extraction}
\label{subsec:openie}

We will now show how $5$ simple hyperedge patterns are sufficient to rank first in a recent Open Information Extraction (OIE) benchmark~\cite{DBLP:conf/acllaw/LechelleGL19}. In fact, one pattern is even sufficient to surpass a majority of the systems of that benchmark. We recognize the limitations of such benchmarks and do not claim that we have the best performing OIE system, neither are we singly focused on this application. Instead, we are interested in providing empirical evidence for the expressive power of SH patterns for the general purpose of knowledge extraction.

To discover OIE patterns, we took advantage of the Wikipedia part of the open text corpus that we developed to train and validate the parser, and that we discussed in section~\ref{sec:text2hyper} -- Wikipedia content is a naturally rich source of factual assertions in NL. The resulting hypergraph contains 62528 top-level hyperedges.


We then employed a simple process of generalization to transform hyperedges into abstract patterns. It consists in replacing each element of a hyperedge with its corresponding type-annotated wildcard, for example:
\Q{(is/P.sc aragorn/C (of/B.ma king/C gondor/C))}
\noindent becomes:
\Q{(*/P.sc */C */C)}

The process can continue recursively, further expanding subedges, for instance: 
\Q{(*/P.sc */C (*/B.ma */C */C))}

These expansions have to conform to Table~\ref{tab:type-inference} (taken in reverse order i.e., from a resulting type to its antecedent), which generally leaves a small number of possibilities.
We further introduced some restrictions in the generation of such patterns, to focus on simple patterns with likely relevance to our task. We limited recursive expansion to depth $2$, and only consider relations of sizes 3 or 4 -- smaller ones cannot contain triplets, larger ones that are useful are likely to contain the triplet (with optional extension) within a core that generalizes to patterns with no more than 4 elements. We excluded conjunctions (these are previously decomposed, as explained in subsection~\ref{sec:conjres}), and modifiers. These latter connectors could certainly be used to improve the OIE task, but entail more semantic complexity, and we are more interested in simplicity at this stage. We will focus on modifiers in a subsequent section. Finally, we allow for the special builder \h{+/B} to be explicitly included in the generalized patterns, given that compound concepts trivially correspond to ontological relationships of OIE interest, e.g.: ``Film director David Lynch'' implies that David Lynch is a film director.

We considered the 50 most common such patterns, which we present in the appendix, in table~\ref{tab:wikipedia-patterns}
. Following the process described in section~\ref{subsec:pattern-learning} and the annotation guidelines document provided with the benchmark~\cite{benchmarkguidelines}, we found that 36 of these patterns can be transformed into valid OIE relationships, given correct parses.

We then compressed these 36 patterns into the most general ones that: (a) imply one or more of the original patterns, and (b) do not imply patterns found to be incorrect in some way. For example, the two patterns:
\Q{(+/B.\{ma\} (ARG1/C...) (ARG2/C...))\\{(+/B.\{mm\} (ARG1/C...) (ARG2/C...))}}
\noindent are compressed to:
\Q{(+/B.\{m[ma]\} */C */C)}

Such a compression/generalization process could feasibly be algorithmically automated.

\begin{table*}[!th]
\centering\small
\begin{tabular}{ c >{\sf}l c c c }
\toprule
 \rm\textbf{\#} & \bf{Pattern} & \textbf{Extractions} & \textbf{F1 {\footnotesize(cumulative)}} & \rm\textbf{Rank} \\
 \midrule
1 & (REL/P.\{[sp][cora]x\} ARG1/C ARG2 ARG3...) & 107 & .265 & 4 \\
\rcg{}2 & (+/B.\{m[ma]\} (ARG1/C...) (ARG2/C...)) & 38 & .311 & 3 \\
3 & (REL1/P.\{sx\}-oc ARG1/C (REL2/T ARG2)) & 20 & .334 & 3 \\
\rcg{}4 & (REL1/P.\{px\} ARG1/C (REL2/T ARG2)) & 12 & .351 & 2 \\
5 & (REL1/P.\{sc\} ARG1/C (REL3/B REL2/C ARG2/C)) & 16 & .365 & 1 \\
 \bottomrule
\end{tabular}
\caption{Open Information Extraction patterns, ordered by decreasing contribution to F1 (presented cumulatively). Ranks correspond to the rank achieved in the benchmark of Table \ref{tab:openie-performance} by using patterns up to the given line.}
\label{tab:openie-patterns}
\end{table*}

We thus arrived at the $5$ patterns which are shown in table~\ref{tab:openie-patterns}. The extracted variables imply the usual OIE tuples: \T{$\langle$REL, ARG1, ARG2, ARG3...$\rangle$}, with argument(s) \T{ARG3...} being optional. Naturally, we convert hyperedges to the actual text they correspond to before feeding them to the benchmark. In the absence of \T{REL}, the relationship ``is'' is assumed. In some cases (patterns 3, 4, 5), notice also that \T{REL} is split into two or thee variables: \T{REL1}, \T{REL2} and \T{REL2}. We just concatenate their textual representation in the order indicated by the variable names, with interleaving space characters.

\medskip



\begin{table*}
\centering\small
\begin{tabular}{ l | c c c | c c | c c c }
\toprule
 \rm\textbf{System} & \rm\textbf{Extractions} & \rm\textbf{Matches}
 & \rm\textbf{Exact} & \rm\textbf{Prec. of} & \rm\textbf{Recall of}
 & \rm\textbf{Prec.} & \rm\textbf{Recall} & \rm\textbf{F1} \\
 & & & \rm\textbf{matches} & \rm\textbf{matches} & \rm\textbf{matches}
 & & & \\
 \midrule
Semantic hypergraphs with 5 rules 
& 201 & 120 & 19 & .70 & \textbf{.93} & .416 & \textbf{.326} & \textbf{.365} \\
MinIE~\cite{gashteovski2017minie} & 252 & \textbf{134} & 10 & .75 & .83 & .400 & .323 & .358 \\
ClausIE~\cite{del2013clausie} & 223 & 121 & \textbf{24} & .74 & .84 & .401 & .298 & .342 \\
OpenIE 4~\cite{mausam2016open} & 101 & 74 & 5 & .68 & .84 & .501 & .182 & .267 \\
Semantic hypergraphs with 1 rule & 107 & 74 & 7 & .69 & .85 & .475 & .184 & .265 \\
Ollie~\cite{mausam2012open} & 145 & 74 & 8 & .73 & .81 & .347 & .175 & .239 \\
ReVerb~\cite{fader2011identifying} & 79 & 54 & 13 & \textbf{.83} & .77 & \textbf{.569} & .121 & .200 \\
Stanford~\cite{angeli2015leveraging} & 371 & 99 & 2 & .79 & .65 & .210 & .188 & .198 \\
PropS~\cite{stanovsky2016getting} & 184 & 69 & 0 & .59 & .80 & .222 & .162 & .187 \\
 \bottomrule
\end{tabular}
\caption{Performance of OpenIE systems, ordered by descending F1. Bold figures indicate the best performing system for each category.}
\label{tab:openie-performance}
\end{table*}

Notice that the first pattern is almost a tautology of the SH representation itself, producing triples where the first argument is the active or passive subject, the relation is the predicate, and the second argument is the direct or indirect object, or complement, or agent, with the optional argument being one of the specifications, if they exist. To illustrate with a real and straightforward example from the benchmark, consider the sentence: ``The population of the special wards is over 9 million people, with the total population of the prefecture exceeding 13 million''. It is parsed to:
\Q{(is/P.scx (of/B.ma (the/M population/C) (the/M (special/M wards/C))) ((over/M (9/M million/M)) people/C) (with/T (exceeding/P.so (of/B.ma (the/M (total/M population/C)) (the/M prefecture/C)) (13/M million/C))))}
\noindent It matches pattern $1$ with variables \T{REL\,=\,is/P.scx}; \T{ARG1\,=\,(of/B.ma (the/M population/C) (the/M (special/M wards/C)))}; \T{ARG2\,=\,((over/M (9/M million/M)) people/C} and \T{ARG3\,=\,(with/T (exceeding/P.so (of/B.ma (the/M (total/M population/C)) (the/M prefecture/C)) (13/M million/C)))}, resulting in the extraction: \emph{$\langle$the population of the special wards, is, over 9 million people, with the total population of the prefecture exceeding 13 million$\rangle$}.

Pattern $2$ can also be seen as a direct consequence of SH representation, in this case inferring ontological relationship from the \h{+/B} builder structure. Cases where both arguments have the role ``\T{m}'' can be interpreted as two expressions of the same concept. Again, using a real example, the emphasized part of the sentence ``He is the younger brother of \emph{the prolific film composer Christophe Beck}'' was parsed as:
\Q{(+/B.mm (the/M (prolific/M (+/B.am film/C composer/C))) (+/B.am christophe/C beck/C))}
\noindent leading to the symmetrical extractions: \emph{$\langle$the prolific film composer, is, Christophe Beck$\rangle$} and \emph{$\langle$Christophe Beck, is, the prolific film composer$\rangle$}. We discuss below in section~\ref{sec:coref} how easy it is to further extract, from this point and thanks to the recursive hypergraphic structure, the relation \emph{$\langle$Christophe Beck, is, film composer$\rangle$}; for now however this is not needed for our comparison with the OIE benchmark.

For a non-symmetrical example with \h{+/B.ma}, let us consider the sentence: ``Finnish police reprimanded a man for traveling in a car boot to hide his meeting with \emph{Prime Minister Juha Sipila} during a government crisis last summer, saying this was breach of the traffic code'', with the emphasized concept parsed as:
\Q{(+/B.am (+/B.am prime/C minister/C) (+/B.am juha/C sipila/C))}
\noindent Using the same pattern, this leads to the single extraction: \emph{$\langle$Juha Sipila, is, Prime Minister$\rangle$}, while avoiding the potentially excessive generalization: \emph{$\langle$Prime Minister, is, Juha Sipila$\rangle$}. We do know that Juha Sipila is one Prime Minister, but not necessarily the only one in the context. The restriction requiring non-atomic edges in the arguments of this pattern is a simple mechanism to avoid too trivial, and potentially silly inferences, such as ``Barack is Obama''.

As a final example, let us illustrate pattern $3$ with the sentence: ``Gonzales graduated from Crescent School in Toronto, Ontario, Canada'', parsed as:
\Q{(graduated/P.sx gonzales/C (from/T (in/B.ma (+/B.am crescent/C school/C) (,/J toronto/C (,/J ontario/C canada/C)))))}
\noindent and resulting in extractions where the first part of REL is extracted from the predicate and the second from the trigger: \emph{$\langle$Gonzales, graduated from, Crescent School in Toronto$\rangle$}. The previously discussed conjunction decomposition process leads to the further extractions: \emph{$\langle$Gonzales, graduated from, Crescent School in Ontario$\rangle$} and \emph{$\langle$Gonzales, graduated from, Crescent School in Canada$\rangle$}.

We stop illustrating how these patterns apply here for the sake of succinctness, but hope to have sufficiently shown how they are generic and straightforward manipulations of the structures enabled by SH.

\medskip
Table~\ref{tab:openie-performance} shows the full benchmark, comparing the performance of our approach with seven other methods. SH outperformed all baseline systems in 24.6\% of the cases that tend to consist of complicated combinations of conjunctions and prepositional phrases. For example, the sentence where the next best system is defeated by the highest margin is: ``A very detailed treatment of the EM method for exponential families was published by Rolf Sundberg in his thesis and several papers following his collaboration with Per Martin-L{\"o}f and Anders Martin-Löf.'' Furthermore, the mean number of words per sentence where SH is not the best is $21.2$, \hbox{vs.} $23.8$ where it is the best ($31.0$ when outperforming by a factor $\ge 1.5$), plausibly indicating an advantage with more complicated sentences.

\section{Computations on the hypergraph: concepts, ontologies and coreference resolution}
\label{sec:corefs}

We have shown how SH representation makes it possible to infer knowledge using simple symbolic rules. We will now address how it enables knowledge inference using probabilistic and heuristic rules. More specifically, we will show how to derive ontologies and perform coreference resolution among concepts. For example, how automated methods can reach the conclusion that ``President Obama'' is a type of ``President'', or that ``Obama'', ``Barack Obama'' and ``President Obama'' refer to the same external entity, while ``Michelle Obama'' refers to another one. Before proceeding, let us consider implicit taxonomies.

\subsection{More about concepts and implicit taxonomies}
\label{subsec:taxonomies}

Hyponyms of a concept can be found by looking for hyperedges where the concept appears either as the main argument of a builder-defined concept or as the argument of a modifier-defined concept.  It follows from these structures that the SH representation implicitly builds a taxonomy.  More generally, we can talk of an implicit ontology. Beyond the taxonomical relationships that we described, the concepts that form a concept hyperedge are related to it in a non-specified fashion. For example, we know that \T{(germany/C)} is related in an unspecified way to \T{(of/B.ma capital/C germany/C)}. Of course, this is not to say that a more specific relation cannot be inferred by further processing with other methods. Here we are simply highlighting the ontological relations that come ``for free'' with the hypergraphic representation.

When parsing sentences to hyperedges, and taking advantage of another classical NLP task offered by the upstream package, we also store auxiliary hyperedges connecting every atom that corresponds to word to the lemma of that word, with the help of a special connector ``\T{lemma/J}''. For instance:
\Q{(lemma/J saying/P say/P)}

In the next section we will make use of this, but it is easy to see how lemmas facilitate the inference of various types of correspondences, for example between singular and plural forms such as \h{apple/C} and \h{apples/C} with the help of \h{lemma/J apple/C apples/C}, and thus more sophisticated structural variations such as \h{+/B.am apple/C season/C} and \h{of/B.ma season/C apples/C}. 

\subsection{Coreference resolution: co-occurrence graph}

\begin{figure*}[!th]
\centering
\includegraphics[width=\linewidth]{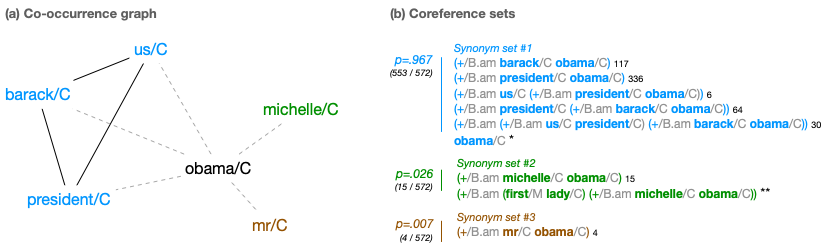}
\caption{Example of coreference resolution. On the left panel we can see the co-occurrence graph and its components, identified by different colors and leading to corresponding coreference sets on the right panel. The probabilities for each coreference set are shown to their left, including the ratios of total degree of the set to total degree, used to compute them. Individual degrees are shown next to each edge.  * indicates the assignment of the seed to one of the coreference sets. ** indicates the recursive nature of the process, with \h{+/B michelle/C obama/C} taking the role of seed in another instance of this coreference resolution process.}
\label{fig:coref}
\end{figure*}

A common but challenging task in NLP is that of coreference resolution, a usual disambiguation issue which consists in identifying different sequences of n-grams that refer to the same entity (such as ``Barack Obama'' and ``President Obama''). This is an old research topic~\cite{soon2001machine} that has been revived lately with modern machine learning methods~\cite{peters2018deep}. ML approaches such as deep learning require large training sets and tend to provide black box models, where precision/recall can be measured and improved upon, but the exact mechanisms by which the models operate remain opaque. Here we do not mean to provide a complete solution to this problem, but instead show that several cases of coreference resolution can be performed in a simpler and understandable manner through the use of semantic hypergraphs for situations that are nevertheless common and useful, especially in the context of social science research.

We will discuss in the following section several experimental results that we obtained on a dataset of several years of news headlines. This corpus is largely focused on political issues, and it is dominated by reports of actors of various types making claims or interacting with each other. These actors can be people, institutions, countries and so on. In our hypergraphic representation, such actors will very frequently be referred to by hyperedges forming compound nouns, with the use of the \T{(+/B)} connector, as discussed previously.

In figure~\ref{fig:coref} we can see such a case: a number of compound concept edges with the main atomic concept \T{(obama/C)} refer to actors. How can we group them in sets, such that all the cases in a given set refer to the same entity? Here, we start taking advantage of the hypergraph as a type of network, and of the analysis graphs that we can easily distill from the hypergraph. Semantic graph-based disambiguation has been extensively explored since the mid-2000s, especially emphasizing the importance of centrality and proximity in deciding which sense correspond to a given word in a certain context, and semantic hypergraphs are no exception \citep{mihalcea2005unsupervised,navigli2007graph,agirre2009personalizing}.

We can trivially traverse all the concepts in the hypergraph, finding the subset of concepts that play the role of main concept in the above mentioned compound concept constructs. For each of these \emph{seed concepts}, we can then attempt to find coreference relationships between the concepts they help build. In the figure, we see an example using the seed concept \T{(obama/C)}. On the right side of the figure, we see all the compound concepts containing the seed as the main concept (except for the ones marked with * and **). It is possible then to build a graph of all the auxiliary concepts that appear together with the seed. A connection between two concepts in this graph means that there is at least a compound concept in which they take part together. In the example, we can see that this graph has three maximal cliques, which we identified with different colors. We then apply this simple rule: two compound concepts are assumed to refer to the same entity if all of their auxiliary concepts belong to the same maximal clique. The intuition is that, if auxiliary concepts belong to a maximal clique, then they tend to be used interchangeably along with the seed, which indicates that they are very likely to refer to the same entity. We will show that this intuition is empirically confirmed in our corpus, from where the example in the figure was extracted.

The co-occurrence graph method produces the coreference sets seen on the right of the figure, except for the items marked with * and **. As can be seen, it correctly groups several variations of hyperedges that refer to Barack Obama (president of the United States during most of the time period covered by our news corpus), and it correctly identifies a separate set referring to Michelle Obama, his wife. It can also be seen that it fails to identify that ``Mr. Obama'' is also likely to refer to Barack Obama. We will say more about this specific case when we discuss claim analysis, in the next section.

But what about the seed concept itself, in this case \T{(obama/C)}? The co-occurrence method is not able to assign it to one of the sets. Here we employ another simple method, this time of a more probabilistic nature. Before tackling this method, we have to make a small detour to discuss the semantic hypergraph from a network analysis perspective.

\subsection{Simple hypergraph metrics}

In a conventional graph, it is common to talk of the degree of a vertex. This refers simply to the number of edges that include this vertex or, in other words, the number of other vertices that it is directly connected to (we assume here an undirected graph without self-loops). With a semantic hypergraph, such measure is not so straightforward, given that an edge can have more than two participants, and that recursivity is permitted.

Let us first define the set $D_e$, containing all the edges in with a given edge $e$ participates:
\begin{equation}
    D_e = \{e_i|e_i \in E \land e \in e_i\}
\end{equation}
We define the degree of a hyperedge $e$ as:
\begin{equation}
    d(e) = \sum_{e_i \in D_e} \big(|e_i| - 1\big)
\end{equation}

This is to say, the hypergraphic degree is the number of edges with which a given edge is connected to by outer hyperedges. It is intuitively equivalent to the conventional graph degree.

Another useful metric that we can define is the \emph{deep degree}, which considers edges connected by hyperedges not necessarily at the same level, but appearing recursively at any level of the connecting hyperedge. Let us consider the set $\Delta_e$, containing the edges that co-participate in other edges with $e$ at any level. This set is recursively defined, so we describe how to generate it, in Algorithm~\ref{algo:generate_delta}.

\SetKwProg{Fn}{Function}{}{end}\SetKwFunction{FRecurs}{Generate$\Delta$}%
\SetAlgoLongEnd

\begin{algorithm}
\DontPrintSemicolon
\Fn(){\FRecurs{e}}{
  \KwData{An edge $e$}
  \KwResult{$\Delta_e$ neighborhood of edge $e$}
  $\Delta_e \longleftarrow D_e$\;
  \For{$e' \in D_e$}{
    $\Delta' \longleftarrow$ \FRecurs{e'} \;
    $\Delta_e \longleftarrow \Delta_e \cup \Delta'$\;
  }
  \KwRet $\Delta_e$ \;
}
\label{algo:generate_delta}
\caption{Generating the neighborhood $\Delta_e$ of an edge $e$.}
\end{algorithm}

\noindent We can now define the deep degree $\delta$ as:
\begin{equation}
    \delta(e) = \sum_{e_i \in \Delta_e} \big(|e_i| - 1\big)
\end{equation}

To provide a more intuitive understanding of these metrics, let us consider the edge ``\T{(is/P berlin/C (of/B capital/C germany/C))}''. Let us also assume that no other edges exist in the hypergraph. In this case, the edges \T{(is/P)}, \T{(of/B capital/C germany/C)} and \T{(germany/C)} all have degree $d=1$, because they all participate exactly in one edge. The first two (\h{is/P} and \h{of/B capital/C germany/C}) also have deep degree $\delta=1$, but the latter \h{germany/C} has deep degree $\delta=2$, because not only does it participate directly in the edge \h{of/B capital/C germany/C}, but it also participates at a deeper level in the outer edge \h{is/P berlin/C (of/B capital/C germany/C)}. In other words, the higher deep degree of \h{germany/C} indicates that it plays an increased role as a \emph{building block} in other edges.

\subsection{Coreference resolution: probabilistic seed assignment}\label{sec:coref}

Back to figure~\ref{fig:coref}, each coreference set is labeled with a probability $p$, representing the chance that a given seed appears in one of its edges, if we were to uniformly enumerate all edges that rely on this seed. This configures a simple estimation of the probability of the seed by itself being used with a certain meaning, represented by the given coreference set. These probabilities are thus the ratio between the sum of the degrees of the edges in each coreference set and the total degree of all edges that include the seed, \hbox{i.e.} of all coreference sets.

Two simple heuristics drive this step. One is that people will tend to use an ambiguous abbreviation of a concept when the popularity of one of the interpretations is sufficiently high in relation to all the others. For example, both  \h{+/B barack/C obama/C} and \h{+/B michelle/C obama/C} share the seed \h{obama/C}, but when referring only to \h{obama/C} during the period he was a US president, people tend to assume that it refers to the most frequently mentioned entity -- \emph{Barack Obama}. The other is that a given seed should only be considered as an abbreviation if it is used a sufficient amount of times as a primary concept in relations, \hbox{i.e.} if there is evidence that it is in fact used on its own to refer directly to some external concept, and not only as a common component of primary concepts. Put differently, seeds referring to common concepts which act often as building blocks of other concepts (\hbox{i.e.}, higher deep degree with respect to degree) are less likely to be valid abbreviations.  Such is the case for ``house'' (which may indifferently refer to the White House or Dr. House) and ``qaida'' (which is typically used as a building block for Al Qaida and never by itself).

We thus establish a criterion that consists of the fulfillment of each of these two conditions, corresponding respectively to the heuristics above. A given seed $s$ is assigned to the coreference set $C$ with the highest $p$ if:

\begin{itemize}
\item  $p$ is above a certain threshold $\theta$
\item ${d_s}/{\delta_s}$ is above a certain threshold $\theta'$
\end{itemize}

We set the threshold to the values $\theta=.7$ and $\theta'=.05$, that we verified empirically to produce good results. Naturally, these thresholds can be fine-tuned using methods more akin to hyper-parameter optimization in ML, but such optimizations are outside of the scope of this work. When the criterion is not met, the seed is left as a reference to a distinct entity. In our corpus, this happens for example with "Korea", which remains an ambiguous reference to either ``North Korea'' or ``South Korea''.

\subsection{Further disambiguation cases}

We do not present here a general solution for coreference resolution and synonym detection, let alone disambiguation as a whole. Some further cases beyond coreference resolution will nonetheless be covered in the next section, notably anaphora resolution, given that this requires the discussion of predicates and relations in more detail, and along with empirical results. Other cases are left out of this work, but we would like to provide a quick insight into how they may be treated.

One obvious example is that of synonyms, which are not implied by a pure structural analysis of hyperedges -- \hbox{e.g.} \emph{red} and \emph{crimson}, as well as \emph{U.S.} and \emph{United States}, for they share no common seed (as opposed to the cases emphasized in the previous subsection). This type of synonym detection may be achieved with the help of a general-knowledge ontology such as Wordnet or DBPedia, and/or with the help of word embeddings such as word2vec. This is a foreseeable and desirable improvement to hypergraph-based text analysis that we leave for future work.

Another case is the inverse problem of synonym detection: disambiguating atoms that correspond to the same word but to different entities, for example distinguishing ``Cambridge (UK)'' from ``Cambridge (USA/Massachusetts)''. We do not perform this type of distinction in this work, but we present another syntactic detail that enables them from a knowledge representation perspective: the atom namespace. Quite simply, beyond the human-readable part and the type and other machine-oriented codes, a third optional slash-separated part can be added to atoms, allowing to distinguish them in cases such as the above, e.g.: \T{cambridge/C/1} and \T{cambridge/C/2}.

Finally, coreference resolution can also apply to cases where neither seed concepts are shared, nor anaphoras are present. Let us say that one sentence refers to ``Kubrick'' and the next one to ``the film director''. Both this type of case and the above mentioned disambiguation cases are likely to be more easily solved with the help of structured knowledge surrounding the concepts in the semantic hypergraph, eventually including general knowledge as mentioned. For example, it could be detected that a certain reference to ``Cambridge'' is closer to references related to the United States, or that ``Kubrick'' is structurally close to the concept of ``film director''. Alternatively, a hybrid approach taking advantage of deep learning models can be employed. In fact, we successfully integrated such a system\footnote{https://github.com/huggingface/neuralcoref} with the Graphbrain library, but avoid using it in this work for the sake of simplicity while defining SH methodological foundations.

\section{Integrated Case Study: Claim and Conflict Analysis}
\label{sec:claim-and-conflict}

We arrive at the point where we can propose an integrated application of the formalisms and methods discussed so far to the analysis of a large corpus of real text, combining symbolic and probabilistic rules. More specifically, we worked with a corpus of news titles that were shared on the social news aggregator \emph{Reddit}. We extracted all titles shared between January 1st, 2013 and August 1st,  2017 on \emph{r/worldnews}, a community that is described as: \emph{``A place for major news from around the world, excluding US-internal news.''} This resulted in a corpus of 404,043 news titles. We applied the methods described in sections~\ref{sec:text2hyper} and \ref{sec:corefs} to generate a hypergraph from the titles. 

We decided to focus on two specific categories of utterances that are very frequent in news sources, and of special interest for the social sciences \cite{tilly-1997-parliamentarization}, especially the study of public spaces \cite{ruiz2016more,van2017clause}: a \emph{claim} made by an actor about some topic and an expression of \emph{conflict} of one actor against another, over some topic. Helpfully, the detection of such categories also allows us to illustrate simple symbolic inference over the hypergraph.

\subsection{Knowledge Inference}

\begin{figure*}
\centering\includegraphics[width=\linewidth]{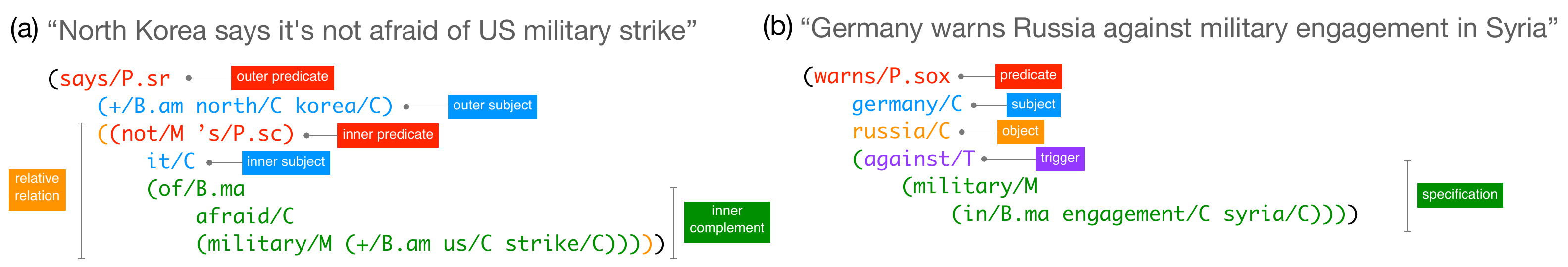}
\caption{\label{fig:patterns}Two examples of relations starting with either a claim or a conflict predicate.}
\end{figure*}

The English language allows for vast numbers of verb constructions that indicate claims or expressions of conflict. Instead of attempting to identify all of them, we considered the $100$ most common predicate lemmas in the hypergraph, and from there we identified a set of ``claim predicates'' and a set of ``conflict predicates'', that we detail below. Overall, we found $3730$ different predicate lemmas, and their rank-frequency distribution is expectedly heavy-tailed. In this case, this small fraction of the set accounts for $60.6\%$ of the hyperedges. Naturally, coverage could be improved by considering more predicates, but with diminishing returns. 

We then employed the same process described in subsection~\ref{subsec:pattern-learning} to discover rules that are capable of detecting claims and expressions of conflict, whereby a hyperedge contains an attributable: 
\begin{itemize}
\item {\bf claim}, if the following conjunction of patterns is satisfied:
\begin{multline*}
\text{\h{PRED/P.\{sr\} \, ACTOR/C \, CLAIM/[RS]}} \\ \land \; \text{\h{lemma/J \, >PRED/P \, [say,claim]/P}}
\end{multline*}

\item {\bf expression of conflict}, if the following conjunction of patterns is satisfied:
{\begin{multline*}
\text{\T{(\,PRED/P.\{so,x\} \, SOURCE/C \, TARGET/C}} \\ \text{\T{[against,for,of,over]/T \,TOPIC/[RS]\,)}} \\
\land \; \text{\T{(\,lemma/J \, >PRED/P}} \\ \text{\T{[accuse,arrest,clash,condemn,kill,slam,warn]/P\,)}}
\end{multline*}}
\end{itemize}

\noindent So, a claim is essentially a relation, based on a predicate of lemma ``say'' or ``claim'', between an actor and a claim, which may also be a relation or a specifier. These patterns additionally rely on a new notation, ``>''. As we have seen in section~\ref{sec:hypergraphs}, a predicate can be a non-atomic hyperedge. As with concepts, the meaning of predicate atoms can be extended with a modifier. For example, the English verb conjugation \emph{``was saying''} is represented as \T{(was/M saying/P)}. Eventually, there is always a predicate atom that corresponds to the main verb in the predicate: 
the notation refers to the innermost atom, i.e. removing an arbitrary amount of nesting based on modifiers. For example, ``\T{>PRED}'' matches ``\T{been/P}'', ``\T{(has/M been/P)}'', ``\T{(not/M (has/M been/P))}'', and so on.

\bigskip In figure~\ref{fig:patterns} we present two real sentences from our corpus and their respective hyperedges. Example (a) was classified as a claim and example (b) as an expression of conflict. These examples 
were purposely chosen to be simple, but the above rules can match more complicated cases. For example, the following sentence was correctly identified and parsed as a claim:
\begin{quote}
U.S. Secretary of State John Kerry was the intended target of rocket strikes in Afghanistan's capital Saturday, the Taliban said in a statement claiming responsibility for the attacks.
\end{quote}

\paragraph{Validation.}
In table~\ref{tab:sentence-parse-validation} we present an evaluation of accuracy based on the manual inspection of $100$ claims and $100$ conflicts that were randomly selected from the hypergraph. Defects are deemed to be minor if they do not interfere with the overall meaning of the hyperedge (e.g., by leading to one of the other, more serious errors listed in the table). To illustrate with a minor defect from our dataset:
\Q{(claims/P.sr (+/B.am google/C boss/C) ((does/M (not/M know/P.so)) he/C (in/B.ma (his/M salary/C) (+/B.am commons/C grilling/C))))}
\noindent In this case, the concept ``commons grilling'' should be a separate specification:
\Q{(claims/P.sr (+/B.am google/C boss/C) ((does/M (not/M know/P.sox)) he/C (his/M salary/C) (in/T (+/B.am commons/C grilling/C))))}

\begin{table}[!t]
\footnotesize\centering
\begin{tabular}{llcc}
\emph{task}&\emph{error type}&\emph{error}&\emph{error}\\
&&\em count&\em rate\\\toprule
claim inference & not a claim & 0/100 & 0\%\\
& wrong actor & 0/100 & 0\%\\
& wrong topic & 2/100 & 2\%\\
& bad anaphora resolution & 1/13 & 8\%\\
& minor defects in topic & 8/100 & 8\%\\\midrule
conflict inference & not a conflict & 0/100 & 0\%\\
& wrong origin actor & 0/100 & 0\%\\
& wrong target actor & 0/100 & 0\%\\
& wrong topic & 0/100 & 0\%\\
& minor defects in topic & 4/100 & 4\%\\
\bottomrule
\end{tabular}
\caption{\label{tab:sentence-parse-validation}Evaluation of several types of error in claim and conflict inference. Error counts and rates are presented, based on the manual inspection of 100 randomly selected claims and 100 randomly selected conflicts.}
\end{table}

\paragraph{Subjects and actors.} Both claim and conflict structures imply that the hyperedge playing the role of subject in the relation is an actor. Using the methods described in section~\ref{sec:corefs}, we can identify the coreference set of each actor and replace all occurrences of this actor with the same hyperedge. For each coreference set we choose the hyperedge with the highest degree as the main identifier, following the heuristic that the most commonly used designation of an entity should be both recognizable and sufficiently compact.

\begin{table*}[!t]
    \centering
    \footnotesize
    \begin{tabularx}{\linewidth}{>{\em}crl>{\sf}Xr}
        \toprule
        \bf{Type} & \bf{Rank} & \bf{Actor} & \bf{Hyperedges (coreference set)} & \bf{Degree}  \\ \midrule
        \multirow{3}{*}{non-human} & 1 & China & china/C, (+/B.am south/C china/C) & 6199 \\
                  & 2 & Russia & russia/C & 5861  \\
                  & 3 & U.S. & us/C, (the/M us/C) & 3824 \\ \midrule
        \multirow{10}{*}{male} & 8 & Vladimir Putin & (+/B.am president/C putin/C), putin/C, (+/B.am vladimir/C putin/C), (+/B.am president/C (+/B.am vladimir/C putin/C)), (+/B.am (russian/M president/C) (+/B.am vladimir/C putin/C)), (+/B.am (russian/M president/C) putin/C) & 2338 \\
             & 10 & Barack Obama & (+/B.am (+/B.am us/C president/C) (+/B.am barack/C obama/C)), (+/B.am president/C obama/C), (+/B.am barack/C obama/C), (+/B.am president/C (+/B.am barack/C obama/C)), obama/C, (+/B.am (+/B.am u.s./C president/C) (+/B.am barack/C obama/C)) & 2069 \\
             & 23 & Donald Trump & (+/B.am president/C (+/B.am donald/C trump/C)), (+/B.am (+/B.am us/C president/C) (+/B.am donald/C trump/C)), (+/B.am donald/C trump/C), trump/C, (+/B.am president/C trump/C) & 1082 \\ \midrule
        \multirow{6}{*}{female} & 32 & Angela Merkel & merkel/C, (+/B.am angela/C merkel/C), (+/B.am (german/M chancellor/C) (+/B.am angela/C merkel/C)), (+/B.am chancellor/C (+/B.am angela/C merkel/C)), (+/B.am (german/M chancellor/C) merkel/C) & 750 \\
               & 78 & Theresa May & may/C, (+/B.am theresa/C may/C), (+/B.am (+/B.am prime/C minister/C) (+/B.am theresa/C may/C)) & 270 \\
               & 201 & Nicola Sturgeon & sturgeon/C, (+/B.am nicola/C sturgeon/C) & 81 \\ \midrule
        \multirow{3}{*}{group} & 46 & The Palestinians & palestinians/C, (the/M palestinians/C) & 487 \\
              & 70 & The Kurds & kurds/C, (the/M kurds/C) & 302  \\
              & 113 & The Russians & russians/C, (the/M russians/C) & 184 \\
        \bottomrule
    \end{tabularx}
    \caption{Three actors with highest hypergraphic degree in each category: non-human, male, female, group  (in decreasing order of highest degree).}
    \label{tab:top_actors}
\end{table*}

As seen in figure~\ref{fig:patterns}(a), the inner subject (i.e., the subject of the relative relation that represents what is being claimed) can be a pronoun. These cases are very common, and almost always correspond to a case where the actor is referencing itself in the content of the claim. On one hand, we perform simple anaphora resolution: if the inner subject is a pronoun in the set \{\T{he/C}, \T{it/C}, \T{she/C}, \T{they/C}\}, then we replace it with the outer subject. On the other hand, we take advantage of the pronoun to infer more things about the actor. The four pronouns mentioned indicate, respectively, that the actor is a male human, a non-human entity, a female human, or a group. We take the majority case, when available, to assign one of these categories to actors. The pronoun \emph{they} is being increasingly used as a gender-neutral third person singular case, but we have not found such cases in our corpus.

Table~\ref{tab:top_actors} shows the top three actors per category, ranked by their degree in the hypergraph, along with their coreference set. Obviously, more sophisticated rules can be devised, both for anaphora resolution and category classification. Our goal here is to illustrate that, thanks to the SH abstraction, it becomes possible to perform powerful inferences (i.e., both useful and at a high level of semantic abstraction) with very simple rules.

\subsection{Topic Structure}
\label{subsec:topic-structure}

The very definition of \emph{topic}, for the purpose of automatic text analysis, is somewhat contingent on the method being employed.  One of the most popular topic detection methods in use nowadays is \emph{Latent Dirichlet Allocation (LDA)}~\cite{blei2003latent}, which is a probabilistic topic model~\cite{blei2012probabilistic} that views topics as latent entities that explain similarities between sets of documents. In LDA, topics are abstract constructs. Documents are seen as a random mixture of topics, and topics are characterized by their probability of generating each of the words found in the document set. LDA uses a generative process to statistically infer these probabilities. Ultimately, a topic is described by the set of words with the highest probabilities of being generated by it. Human observers can then infer some higher-level concept from these words and probabilities. For example, if the five highest probability words for a topic $X$ are \emph{\{EU, Ursula von der Leyen, Boris Johnson, Barnier, Trade\}}, a human observer may guess that a good label for this topic is \emph{Brexit Negotiations}. LDA is applicable to sets of documents, for a predefined number of topics, where each document is considered to be a \emph{bag-of-words}. Distributional topic detection methods have recently generated a variety of research endeavors, including the application of stochastic blockmodels to discover joint groups of documents which use keywords in a similar fashion \citep{gerlach2018network}.

A different approach to topic detection is \emph{TextRank}~\cite{mihalcea2004textrank}, which is capable of detecting topics within a single document. With TextRank, the document is first transformed into a word co-occurrence graph. Common NLP approaches are used to filter out certain classes words from the graph (e.g., do not consider articles such as ``the''). Topics are considered to be the words with the highest network centrality in this graph, according to some predefined threshold condition. Simple statistical methods over the co-occurrence graph can be used to derive \emph{ngram} topics from the previous step. Given that the order in which words appear in the document is important, TextRank cannot be said to be a \emph{bag-of-words} approach such as LDA. It relies a bit more on the meaning of the text, and it is more local -- in the sense that it works inside a single document instead of requiring statistical analysis over a corpus of documents.

\begin{table*}[t]
\scriptsize
\begin{tabularx}{\linewidth}{>{\bf\scriptsize}lp{2.3cm}ccX}
&\scriptsize\emph{actor}&&&\scriptsize\emph{topic}\\\toprule
&&&$\rightarrow$&scuppering syria peace talks\\\multirow{-2}{*}{$\Longrightarrow$}&\multirow{-2}{*}{assad}&&$\rightarrow$&war crimes in aleppo\\\rowcolor{gray!10}\multirow{-1}{*}{$\Longrightarrow$}&\multirow{-1}{*}{damascus}&&$\rightarrow$&continuing to use chemical weapons\\\multirow{-1}{*}{$\Longrightarrow$}&\multirow{-1}{*}{the united states}&&$\rightarrow$&the weakening of europe\\\rowcolor{gray!10}\multirow{-1}{*}{$\Longrightarrow$}&\multirow{-1}{*}{us}&&$\rightarrow$&espionage\\&&$\leftarrow$&&mistral delay\\&&&$\rightarrow$&meddling in election\\\multirow{-3}{*}{$\Longleftrightarrow$}&\multirow{-3}{*}{russia}&&$\rightarrow$&rapid eu sanctions\\\rowcolor{gray!10}&&$\leftarrow$&&being europe's biggest problem child\\\rowcolor{gray!10}\multirow{-2}{*}{$\Longleftrightarrow$}&\multirow{-2}{*}{germany}&&$\rightarrow$&wage dumping in the meat sector\\\multirow{-1}{*}{$\Longleftarrow$}&\multirow{-1}{*}{al qaeda}&$\leftarrow$&&new attacks\\\rowcolor{gray!10}\multirow{-1}{*}{$\Longleftarrow$}&\multirow{-1}{*}{council of europe}&$\leftarrow$&&allowing to hit parents and spank their children\\\multirow{-1}{*}{$\Longleftarrow$}&\multirow{-1}{*}{iraqis}&$\leftarrow$&&imminent isis attacks\\\rowcolor{gray!10}\multirow{-1}{*}{$\Longleftarrow$}&\multirow{-1}{*}{kagame}&$\leftarrow$&&rwanda genocide\\\multirow{-1}{*}{$\Longleftarrow$}&\multirow{-1}{*}{london}&$\leftarrow$&&plot\\\rowcolor{gray!10}\multirow{-1}{*}{$\Longleftarrow$}&\multirow{-1}{*}{syria's assad}&$\leftarrow$&&supporting terrorism\\\multirow{-1}{*}{$\Longleftarrow$}&\multirow{-1}{*}{un}&$\leftarrow$&&racist attacks on black minister\\\rowcolor{gray!10}\multirow{-1}{*}{$\Longleftarrow$}&
{united nations top human rights official}&$\leftarrow$&&delays\\
\bottomrule
\end{tabularx}
\caption{\label{tab-ego-centered-france}List of actors criticizing or being criticized by ego (here, France), and the topics over which the critique applies. Single arrows show the critique direction (left to right: ego criticizes that actor) for each underlying hyperedge, double arrows indicate the overall critique direction (which can thus go both ways).}
\end{table*}

In this work, we move significantly more in the direction of text understanding and locality. Our topics are firstly inferred from the meaning of sentences. As we have shown, pattern analysis of hyperedges can be used to infer relationships such as \emph{claim} and \emph{conflict}, which imply both actors and topics. Given coreference detection, such topics are characterized by sets of hyperedges, but these sets are not probabilistic in the sense that LDA's are. Instead, they are a best guess of symbolic representations that map to some unique concept. Our approach relies even more on meaning than TextRank, and it allows for topic detection at an even more local scale: single sentences.

In the examples given in figure~\ref{fig:patterns}, the claim shown in (a) implies the rather specific topic ``afraid of military us strike'', and (b) the topic ``military engagement in syria''.

Another important aspect of our approach is that topics can be composed of other topics or concepts, forming a hierarchical structure. This is a natural consequence of how we model language, as explained in section~\ref{subsec:taxonomies}. This allows us to explore topics at different levels of detail. The topic implied by a claim or conflict can be very specific and possibly unique in the dataset, but the more general subtopics or concepts that it contains can be used to find commonalities across the hypergraph. Considering the hyperedge from one of the topic examples above, from \h{military/M (in/B engagement/C syria/C)} it is possible to extract concepts from inner edges that correspond to more general concepts, for example \h{syria/C}, \h{engagement/C} and \h{in/B engagement/C syria/C}. With the help of the implicit taxonomy, which indicates that \T{(in/B engagement/C syria/C)} is a type of \T{engagement/C}, a simple rule could also infer that \T{(military/M (in/B engagement/C syria/C))} is a type of \T{(in/B engagement/C syria/C)}. 

In the various tables of results that we will subsequently present, actors and topics are represented by labels in natural language. During the transformation of text to hyperedge, every hyperedge that is generated is associated with the chunks of text from which it comes. These chunks are then used as textual labels for the hyperedges.

\subsection{Inter-actor criticism}

Focusing on France and Germany 
as target actors $a$, we gather the results for the detection of conflict patterns in the tables \ref{tab-ego-centered-france} and \ref{tab-ego-centered-germany}. 
Each of these actors is involved in active or passive criticism of other actors, \hbox{i.e.} either critical (\ra) of or criticized by ($\leftarrow$) other actors.  The critique is related to a topic, and may go in both directions, \hbox{i.e.} Germany criticizes Greece for debt commitments (second row of table~\ref{tab-ego-centered-germany}).

The topics presented here correspond to the detailed topics discussed in the previous section. This structured enumeration provides a way to scan the direction, target and frequency of claims by actors on other actors in a given text corpus.

\begin{table*}[!t]
\footnotesize
\begin{tabularx}{\linewidth}{>{\bf\scriptsize}lp{2.3cm}ccX}
&\scriptsize\emph{actor}&&&\scriptsize\emph{topic}\\\toprule
\multirow{-1}{*}{$\Longrightarrow$}&\multirow{-1}{*}{fiat}&&$\rightarrow$&using illegal emissions device\\\rowcolor{gray!10}\multirow{-1}{*}{$\Longrightarrow$}&\multirow{-1}{*}{greece}&&$\rightarrow$&debt commitments\\\multirow{-1}{*}{$\Longrightarrow$}&\multirow{-1}{*}{israel}&&$\rightarrow$&latest settlement expansion in east jerusalem\\\rowcolor{gray!10}\multirow{-1}{*}{$\Longrightarrow$}&\multirow{-1}{*}{kurds}&&$\rightarrow$&one sided referendum plans\\\multirow{-1}{*}{$\Longrightarrow$}&\multirow{-1}{*}{maduro}&&$\rightarrow$&holding venezuelans' hostage\\\rowcolor{gray!10}\multirow{-1}{*}{$\Longrightarrow$}&\multirow{-1}{*}{mexico city}&&$\rightarrow$&brexit\\&&&$\rightarrow$&cold war reflexes\\\multirow{-2}{*}{$\Longrightarrow$}&\multirow{-2}{*}{putin}&&$\rightarrow$&moscow up beefs nuclear arsenal\\\rowcolor{gray!10}\multirow{-1}{*}{$\Longrightarrow$}&\multirow{-1}{*}{syrian}&&$\rightarrow$&alleged car bomb plot\\\multirow{-1}{*}{$\Longrightarrow$}&\multirow{-1}{*}{uk}&&$\rightarrow$&leaving eu\\\rowcolor{gray!10}\multirow{-1}{*}{$\Longrightarrow$}&\multirow{-1}{*}{ukraine}&&$\rightarrow$&graft\\\multirow{-1}{*}{$\Longrightarrow$}&\multirow{-1}{*}{us}&&$\rightarrow$&stasi methods ahead of obama\\\rowcolor{gray!10}&&$\leftarrow$&&halting arms deal\\\rowcolor{gray!10}&&&$\rightarrow$&cyber attack on ukraine peace monitors\\\rowcolor{gray!10}&&&$\rightarrow$&kremlin dismisses us intelligence claims as a witch hunt\\\rowcolor{gray!10}\multirow{-4}{*}{$\Longleftrightarrow$}&\multirow{-4}{*}{russia}&&$\rightarrow$&military engagement in syria\\&&$\leftarrow$&&wage dumping in the meat sector\\\multirow{-2}{*}{$\Longleftrightarrow$}&\multirow{-2}{*}{france}&&$\rightarrow$&being europe's biggest problem child\\\rowcolor{gray!10}&&$\leftarrow$&&causing instability\\\rowcolor{gray!10}\multirow{-2}{*}{$\Longleftrightarrow$}&\multirow{-2}{*}{u.s.}&&$\rightarrow$&ceding lead role to china\\&&$\leftarrow$&&backing failed coup\\&&$\leftarrow$&&cultural racism over eu accession\\&&$\leftarrow$&&engaging in diplomatic rudeness and double standards\\&&$\leftarrow$&&genocide speech\\&&$\leftarrow$&&harbouring terrorists\\&&$\leftarrow$&&succor providing to its enemies\\&&$\leftarrow$&&succour providing to its enemies\\&&$\leftarrow$&&working against erdogan\\&&&$\rightarrow$&blackmailing eu\\&&&$\rightarrow$&itself further distancing from europe by the death penalty reinstating after a disputed referendum\\&&&$\rightarrow$&monday\\&&&$\rightarrow$&nazi\\\multirow{-13}{*}{$\Longleftrightarrow$}&\multirow{-13}{*}{turkey}&&$\rightarrow$&supporting terrorism\\\rowcolor{gray!10}\multirow{-1}{*}{$\Longleftarrow$}&\multirow{-1}{*}{erdogan}&$\leftarrow$&&nazi practices over blocked political rallies\\\multirow{-1}{*}{$\Longleftarrow$}&\multirow{-1}{*}{eu commission}&$\leftarrow$&&air pollution breaches\\\rowcolor{gray!10}\multirow{-1}{*}{$\Longleftarrow$}&\multirow{-1}{*}{eu leaders}&$\leftarrow$&&pressure on migrant quotas\\\multirow{-1}{*}{$\Longleftarrow$}&
{french far right leader marine le pen}&$\leftarrow$&&doors opening to refugees\\\rowcolor{gray!10}\multirow{-1}{*}{$\Longleftarrow$}&\multirow{-1}{*}{italy}&$\leftarrow$&&undermining its economic efforts\\\multirow{-1}{*}{$\Longleftarrow$}&\multirow{-1}{*}{moscow}&$\leftarrow$&&up hushing russian girl's rape\\\rowcolor{gray!10}\multirow{-1}{*}{$\Longleftarrow$}&\multirow{-1}{*}{orban}&$\leftarrow$&&rude tone over refugees\\\multirow{-1}{*}{$\Longleftarrow$}&\multirow{-1}{*}{snowden}&$\leftarrow$&&nsa aiding in spying efforts\\\rowcolor{gray!10}\multirow{-1}{*}{$\Longleftarrow$}&
{turkey's president tayyip erdogan}&$\leftarrow$&&behaving like nazis\\\multirow{-1}{*}{$\Longleftarrow$}&\multirow{-1}{*}{un}&$\leftarrow$&&institutional racism and racist stereotyping against people of african descent\\\rowcolor{gray!10}\multirow{-1}{*}{$\Longleftarrow$}&\multirow{-1}{*}{un committee}&$\leftarrow$&&an anti racism - convention violating by not prosecuting a politician's comments about turks and arabs\\
\bottomrule
\end{tabularx}
\caption{\label{tab-ego-centered-germany}List of actors criticizing or being criticized by ego (here, Germany), and the topics over which the critique applies. Single arrows show the critique direction (left to right: ego criticizes that actor) for each underlying hyperedge, double arrows indicate the overall critique direction (which can thus go both ways).} 
\end{table*}

\subsection{Dyadic claims}

Here we focus on claims that actors make about other actors (or themselves). In other words, we refer to claims where the subject of the claim is itself an actor. Furthermore, we consider only claims for which the claim relation contains an argument playing the rule of complement, meaning that the subject of the claim is being linked with some concept, for example expressing membership in a class (e.g.: ``Pablo is a cat.'') or the possession of some property (e.g.: ``North Korea is afraid'').

We also recursively extract context edges that are connected to the outer claim edge through nestings of \h{:/J}. To give an example from our corpus:
\Q{(:/J (says/P.sr russia/C ('s/P.sc it/C ready/C)) ((to/M deal/P.x) (with/T (new/M (+/B.am ukraine/C president/C)))))}

The edge ``\T{((to/M deal/P.x) (with/T (new/M (+/B.am ukraine/C president/C))))}'' is extracted as a context edge. Finally, specification edges of the claim and context edges are extracted out and grouped together.

The predicate of the relative relation that expressed the claim is inspected to further determine the tense of the attribution (present, past, future
), and to identify negations. Once again, this is achieved by simple rules over the hypergraphic representation:

\begin{itemize}
\item The presence of a negation modifier ({\T{not/M}, \T{n't/M}}), as is in fact the case with the first example of figure~\ref{fig:patterns}.
\item The presence of the predicate {\T{was/P}} implies the past.
\item The presence of the modifier {\T{will/M}} implies the future tense.
\end{itemize}

\begin{table*}[ht]
\scriptsize\centering
\begin{tabularx}{\linewidth}{c>{\columncolor[gray]{0.97}\raggedright}cc>{\columncolor[gray]{0.97}\raggedright}clX}
\textbf{source}&\multicolumn{1}{c}{}&\textbf{target}&\multicolumn{1}{c}{}&\multicolumn{2}{l}{\textbf{property, \emph{context} and \emph{<specification>}}}\\\toprule
    &   &   &   & &just the place for you\\
    &   &   &   & &the victim of intensive cyberattacks\\
    &   &   &   & &ready; \emph{to strike u.s. aircraft carrier}\\
    &   &   &   & &able; \emph{to nuke u.s. mainland}\\
    &   &   &   & &a great place for human rights\\
    &   &   &   & &ready for war with us\\
    &   &   &   & &close; \emph{developing a new satellite}; \emph{speculation fuelling it might attempt a long range rocket to fire to mark a key political anniversary}; \emph{<next month>}\\
    &   &   &   & &open; \emph{holding talks with south korea}; \emph{<including the suspension of the south's joint military drills with the united states>}; \emph{<if are met certain conditions>}\\
    &   &   &   & &open; \emph{to talk with south korea}\\
    &   &   &   & &the biggest victim in u.s. student's death\\\cline{6-6}
    &   &   &   &   &responsible for righteous sony hacking\\
\multirow{-14}{*}{\textbf{north korea}}&\multirow{-14}{*}{says}&\multirow{-14}{*}{\textbf{north korea}}&\multirow{-14}{*}{is}&\multirow{-2}{*}{\emph{not}}&afraid of us military strike\\
\midrule
    &   &   &\cellcolor{white}was & &ready; \emph{to put russia's nuclear weapons}; \emph{<during tensions over the crisis in ukraine and crimea>}; \emph{<on standby>}\\\cline{5-6}
    &   &   &   & &possible convinced solution to ukraine crisis\\
    &   &   &   & &willing; \emph{to play a mediating role between the two koreas}; \emph{relieve the state of crisis on the korean peninsula}; \emph{<according to president moon jae in's special envoy to moscow>}; \emph{<to help>}; \emph{<by dispatching an emissary>}; \emph{<to pyongyang>}\\
    &   &   &   & &ready; \emph{to sell s-400 anti aircraft system}; \emph{<to turkey>}\\\cline{6-6}
    &   &   &\multirow{-6}{*}{is}&\emph{not} &russia's president for life\\\cline{5-6}
    &   &   &\cellcolor{white}& &russia's president; \emph{2024}\\\cline{6-6}
\multirow{-10}{*}{\textbf{putin}}& &    &\multirow{-2}{*}{\cellcolor{white}will be}&\emph{not} &president for life\\\cline{1-1}\cline{5-6}
    &   &   &   & &ready; \emph{to improve ties with the us}\\
    &   &   &   & &moral compass of the world\\
    &   &   &   & &interested; \emph{other brics brazil, russia, india, china, south africa members}; \emph{using national currencies}; \emph{<after agreeing on such an arrangement with china>}\\
    &   &   &   & &willing; \emph{over to hand to us house of representatives and senate}\\\cline{6-6}
\multirow{-6}{*}{\textbf{russia}}&   &  &\multirow{-6}{*}{is}&\emph{not}  &a threat to anyone\\\cline{1-1}\cline{5-6}
    &   &   &\cellcolor{white}was& &the mastermind of the ukrainian coup\\\cline{5-6}
\multirow{-2}{*}{\textbf{us}}&\multirow{-16}{*}{says} &\multirow{-16}{*}{\textbf{putin}}&is & &world's only superpower; \emph{back walks trump compliments}\\
\midrule
    &   &   &   & &open; \emph{coordinating with u.s. in syria}\\
    &   &   &   & &ready; \emph{to deal with new ukraine president}; \emph{to retaliate for u.s. election sanctions}\\
    &   &   &   &   &ready for dialogue with petro poroshenko, ukraine's next president\\
    &   &   &   &   &ready; \emph{to provide the free syrian army}; \emph{<with air support in fight against islamic state>}\\
\multirow{-5}{*}{\textbf{russia}}&\multirow{-5}{*}{says}&\multirow{-5}{*}{\textbf{russia}} &\multirow{-5}{*}{is}& &building naval bases in asia, latin america
\\
\midrule
    &   &   &   & &disappointed over china's failure; \emph{over to hand fugitive intelligence analyst edward snowden}\\
    &   &   &   & &open; \emph{to work with new iranian president hussain rowhani}\\\cline{6-6}
\multirow{-3}{*}{\textbf{us}} & \multirow{-3}{*}{says} &\multirow{-3}{*}{\textbf{us}} &\multirow{-3}{*}{is}&\emph{not}&surprised; \emph{<if north korea launches missiles>}\\
\bottomrule
\end{tabularx}
\caption{\label{tab:dyadic-claims}List of claims, or attributions by subject actors (sources) about other actors (targets). Automatically identified negative claims are emphasized.}
\end{table*}


In table~\ref{tab:dyadic-claims}, we present such attributions between the actors: North Korea, Russia, Putin
and U.S.

\subsection{Topic-based conflict network}

\begin{figure*}
\begin{center}\includegraphics[width=.8\linewidth]{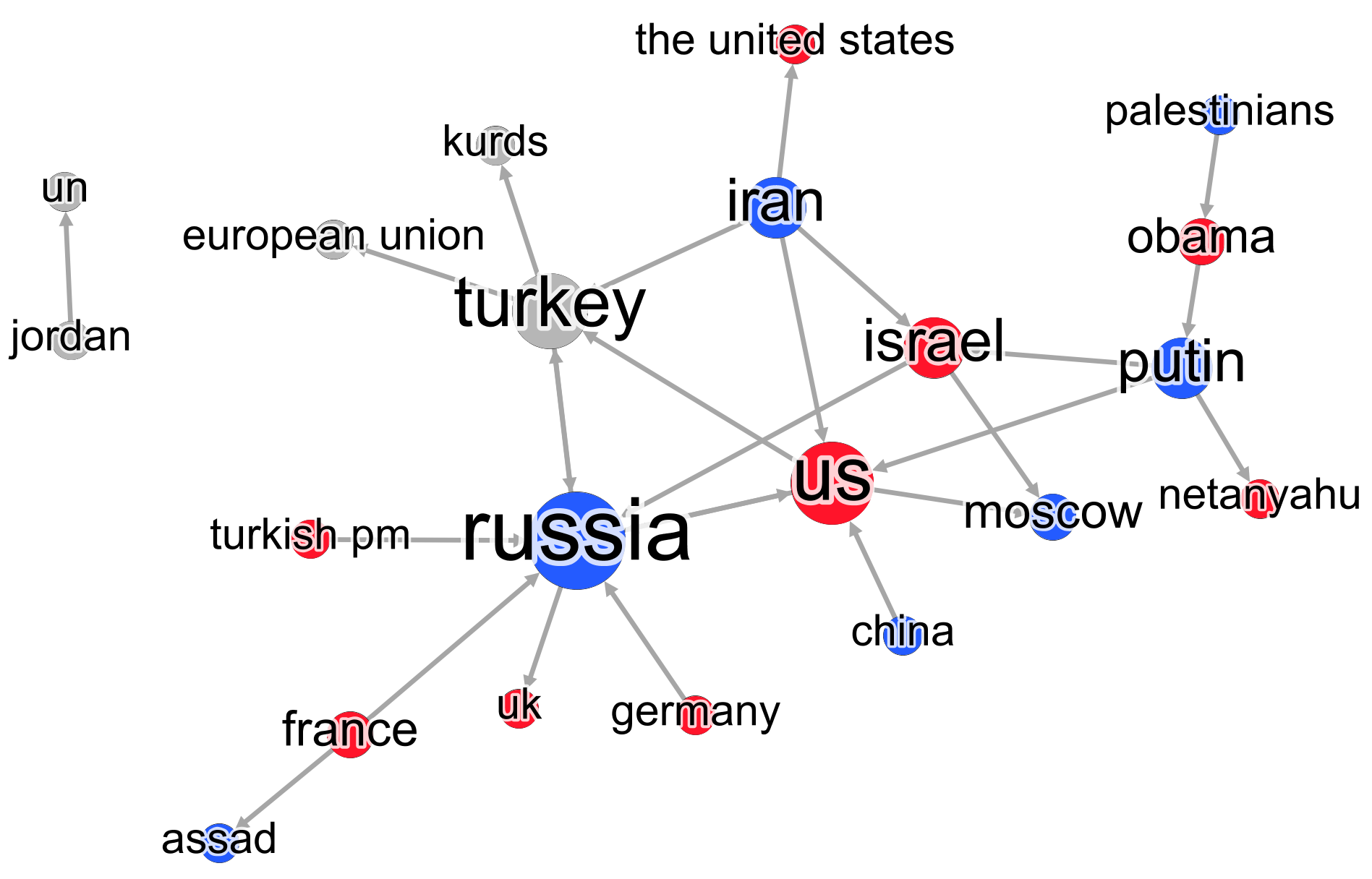}\end{center}
\caption{\label{fig:conflict-net}Network of conflicts between actors over the topic ``Syria''. Arrows point from the originator of the conflict to its target. Size of nodes is proportional to their degree. Two factions were identified by a simple algorithm. One faction is represented as red, the other blue. Gray nodes do not belong to any faction.}
\end{figure*}

So far we have presented actor-centric results. Here we will consider all conflicts that contain ``Syria'' as topic or subtopic (according to the definitions of section~\ref{subsec:topic-structure}). From this set of hyperedges we extracted a directed network connecting actors engaging in expressions of conflict over Syria. A visualization of this network is presented in figure~\ref{fig:conflict-net}.

We devised a very simple algorithm to identify two factions in this conflict graph. Firstly, we attribute a score $s_{ij}$ to every hyperedge ($e_{ij}$):
$$s_{ij} = \min(d_i, d_j)$$
where $d_i$ is the degree (in- and out-) of node $i$. Then we iterate through hyperedges in descending order of $s$. This heuristic assumes that hyperedges connecting more active nodes are more likely to represent the fundamental dividing lines of the overall conflict. The first hyperedge assigns one node to faction A and another to faction B. From then on, a node is assigned to a faction if it does not have a conflict with any current member of this faction, and has a conflict with a current member of the opposite faction. In the case that the node cannot be assigned to any faction, it remains unassigned.

This resulted in faction A containing the actors: \T{\{russia, iran, moscow, putin, china, erdogan, palestinians\}}, faction B the actors: \T{\{us, west, israel, the united states, france, netanyahu, uk, germany, obama\}} and the following actors remaining unassigned: \T{\{turkey, assad, the european union\}}. Faction A is shown in blue in figure~\ref{fig:conflict-net}, and faction B in red. This categorization and network visualization suggest that the main axis of the conflict around Syria is a Russia / U.S conflict. Factions A and B contain state actors and political leaders that are typically aligned with, respectively, Russia and the U.S.

Naturally, more sophisticated faction and alliance detection methods can be employed. Here we are mostly interested in showing the effectiveness of our approach in summarizing complex situations from large natural language corpora, and to provide some empirical validation that these results are sensible. This conflict graph was built from a total of $53$ hyperedges, and we manually verified that they all correspond to expressions of conflict, and that both the intervening actors and main topic were correctly identified in all cases.

\section{Conclusions}

We have presented the novel SH formalism, aimed at a new approach to language understanding based on the idea of translating NL into a structured and formal representation, while allowing room for the inherent and often irreducible ambiguities found in human communication. Developing the formalism entailed an effort of modeling of NL into a set of types and syntactic rules that (1) preserves the richness of NL, (2) facilitates computational language understanding tasks and (3) can act as a \emph{lingua franca} for hybrid systems that include both human and computational intelligence of various natures (e.g. symbolic, graph-based and statistical).

Surrounding this central idea, we presented a viable parser of NL to SH using standard contemporary ML techniques as well as higher-level linguistic features provided by standards NLP libraries. Furthermore, we have shown that inference rules and knowledge extraction patterns can be represented in SH notation itself, and we have developed a procedural template for systems capable of learning inference rules with the collaboration of humans and reference hypergraphs extracted from open text.

We believe to have empirically validated our approach from several angles and in several ways: in terms of the completeness of the representation by its ability to represent all grammatical constructs found in the Universal Dependencies; in terms of the precision of the parser across of variety of text categories; in terms of the expressive power of SH in being able to produce competitive results in a task for which a number of dedicated competing systems and an external benchmark exists; and finally in its ability to tackle a set of specific and related tasks of language understanding of particular interest to the social sciences (actor and gender detection, co-reference resolution, claim and conflict analysis), producing results that reasonably match common-sense intuitions about the ground truth, and also display good precision when manually verified.

The central goal has been to lay the foundations of this approach and demonstrate its potential. We do not claim to have the best performing system in any of the tasks that we tackled, nor was this our aim, but we do hope to have demonstrated the versatility, completeness and potential of SH.

Our further ambition is to apply this method to a variety of language understanding tasks where the understandability of results is desirable, for example in news / social media analysis (detection of actors, topics, conflicts, agreements, causality, beliefs), or to extract viewpoints from scientific articles, or to help in the study of cultural objects such as literary works (known as \emph{Distant Reading}~\cite{moretti2013distant}). Our hope is that we were able to convince the reader of the potential of SH for such tasks, enabling large-scale text corpus analysis while preserving the rich understanding expected in social science endeavors which normally require a significant amount of tedious manual coding \cite{tilly-1997-parliamentarization}.

We created the Graphbrain open-source software library that implements all the ideas we described\footnote{https://github.com/graphbrain/graphbrain}, aiming not only to facilitate the replicability of all the experiments that we performed in this work, but also to facilitate the adoption and extension of SH and related methodology by the research community at large.

\section*{Acknowledgements}

This research has been supported by the ``Socsemics'' Consolidator grant funded by the European Research Council (ERC) under the European Union Horizon 2020 research and innovation program, grant agreement No. 772743.


\appendix

\section{Mapping Universal Stanford Dependencies to hyperedges}
\label{app:universal-deps}

We used the \emph{Universal Stanford Dependencies}~\cite{nivre2016universal} to guide the development of the \emph{Semantic Hypergraphs} representation. In table~\ref{tab:universal}, we present one example of hyperedge for each grammatical relation in the Universal Dependencies. This is meant as empirical evidence for the completeness of our model, in terms of its ability to map to natural language constructs in most human languages.

\begin{table*}
\begin{center}
\small{
\begin{tabular}{ l >{\sf}l }
 \toprule
 \textbf{Grammatical Relation} & \bf{Hyperedge Example}\\
 \midrule

 \textbf{nsubj}: nominal subject & (is/P.sc (the/M \x{dog/C}) cute/C) \\
 \rcg{}\textbf{nsubjpass}: passive nominal subject & ((was/M played/P.pa) (the/M \x{piano/C}) (by/T mary/C)) \\  
 \rcg{}\textbf{dobj}: direct object (accusative) & (gave/P.sio mary/C john/C (a/M \x{gift/C})) \\
 \textbf{iobj}: indirect object (dative) & (gave/P.sio mary/C \x{john/C} (a/M gift/C)) \\ 
 \rcg{}\textbf{csubj}: clausal subject & (makes/P.so (\x{said/P.os} what/C she/C) sense/C) \\
 \textbf{csubjpass}: clausal passive subject & ((was/M suspected/P.pa) (that/T (\x{lied/P.s} she/C)) (by/T everyone/C)) \\
 \rcg{}\textbf{ccomp}: clausal complement & (says/P.so he/C (\x{like/P.so} you/C (to/M swim/C))) \\
 \textbf{xcomp}: open clausal complement & (says/P.so he/C (like/P.so you/C (to/M \x{swim/C}))) \\
 \rcg{}\textbf{nmod}: nominal modifier & ((of/B \x{some/C} (the/M toys/C)) \\
 \textbf{advcl}: adverbial clause modifier & (talked/P.sxx he/C (to/T him/C) (to/T (\x{secure/P.o} (the/M account/C)))) \\
 \rcg{}\textbf{advmod}: adverb modifier & ((\x{genetically/M} modified/M) food/C) \\
 \textbf{neg}: negation modifier & ((\x{not/M} is/P.sc) bill/C (a/M scientist/C)) \\
 \rcg{}\textbf{vocative}: vocative & (know/P.sv i/C \x{john/C}) \\
 \textbf{discourse}: discourse & N/A \\
 \rcg{}\textbf{expl}: expletive & ((\x{there/M} is/P) (a/M (in/B ghost/C (the/M room/C)))) \\
 \textbf{aux}: auxiliary & ((\x{has/M} (been/M killed/P.p)) kennedy/C) \\
 \rcg{}\textbf{auxpass}: passive auxiliary & ((has/M (\x{been/M} killed/P.p)) kennedy/C) \\
 \textbf{cop}: copula & (is/P.sc bill/C \x{big/C}) \\
 \rcg{}\textbf{mark}: marker & (says/P.so he/C (\x{that/T} (like/P.so you/C (to/M swim/C)))) \\
 \textbf{punct}: punctuation & \emph{N/A} \\
 \rcg{}\textbf{conj}: conjunction & (is/P.so bill/C (and/J big/C \x{honest/C})) \\
 \textbf{cc}: coordination & (is/P.so bill/C (\x{and/J} big/C honest/C)) \\
 \rcg{}\textbf{nummod}: numeric modifier & (ate/P.so sam/C (\x{3/M} sheep/C)) \\
 \textbf{relcl}: relative clause modifier & (saw/P.so i/C (the/M (\x{love/P.so} you/C man/C))) \\
 \rcg{}\textbf{det}: determiner & (is/P.sc (\x{the/M} man/C) here/C) \\
 \textbf{compound}: compound & (has/P.so john/C (the/M (+/B phone/C \x{book/C}))) \\
 \rcg{}\textbf{name}: multi-word proper nouns & (+/B \x{marie/C} \x{curie/C}) \\
 \textbf{mwe}: multi-word expression & (cried/P.sx he/C (\x{because/T} (of/M you/C))) \\
 \rcg{}\textbf{foreign}: foreign words & \emph{misc.} \\
 \textbf{goeswith}: goes with & (come/P.sox they/C here/C (with\x{out}/T permission/C)) \\
 \rcg{}\textbf{case}: case marking & (the/M (\x{'s/B} school/C grounds/C)) \\
 \textbf{list}: list & ('s/B mary (:/J list/C (:/B \x{phone/C} 555-981/C) (:/B \x{age/C} 33/C))) \\
 \rcg{}\textbf{dislocated}: dislocated elements & (is/P.scd this/C (our/M office/C) (and/J me/C sam/C)) \\
 \textbf{parataxis}: parataxis & (left/P.tsi (\x{said/P.s} john/C) (the/M guy/C) (in/B early/C (the/M morning/C))) \\
 \rcg{}\textbf{remnant}: remnant in ellipsis & (won/P.sor john/C bronze/C (:/B \x{mary/C} \x{silver/C})) \\
 \textbf{reparandum}: overridden disfluency & (go/P.eo (to/T (the/M \x{righ-/C})) (to/T (the/M left/C))) \\
 \rcg{}\textbf{root}: sentence head & (\x{is/P.sc} (the/M dog/C) cute/C) \\
 \textbf{dep}: unspecified dependency & \emph{misc.} \\
 \bottomrule
\end{tabular}
}
\end{center}
\caption{Examples of hyperedges for each grammatical relation in the Universal Dependencies.}
\label{tab:universal}
\end{table*}

\section{Most common hyperedge patterns}
\label{app:wikipedia-patterns}

Table~\ref{tab:wikipedia-patterns} lists the 50 most commons hyperedge patterns in a hypergraph built from a Wikipedia sample.

\begin{table*}
\begin{center}
\small{
\begin{tabular}{ r >{\sf}l r c }
 \toprule
 \textbf{\#} & \bf{Pattern} & \textbf{\# Cases} & \textbf{OIE Pattern} \\
 \midrule
1 & (*/B.\{ma\} */C */C) & 32396 & - \\
\rcg{}2 & (+/B.\{ma\} */C */C) & 13602 & 2 \\
3 & (*/T */C) & 12536 & - \\
\rcg{}4 & (*/B.\{mm\} */C */C) & 5331 & - \\
5 & (+/B.\{mm\} */C */C) & 5331 & 2 \\
\rcg{}6 & (*/T */R) & 2796 & - \\
7 & (*/P.\{so\} */C */C) & 1572 & 1 \\
\rcg{}8 & (*/P.\{sx\} */C */S) & 964 & 3 \\
9 & (*/P.\{sc\} */C */C) & 916 & 1 \\
\rcg{}10 & (*/P.\{sox\} */C */C */S) & 861 & 1 \\
11 & (*/P.\{ox\} */C */S) & 631 & - \\
\rcg{}12 & (*/P.\{px\} */C */S) & 604 & 4 \\
13 & (*/P.\{sr\} */C */R) & 557 & 1 \\
\rcg{}14 & (*/P.\{sox\} */C (*/B.\{ma\} */C */C) */S) & 337 & 1 \\
15 & (*/P.\{scx\} */C */C */S) & 322 & 1 \\
\rcg{}16 & (*/P.\{so\} (*/B.\{ma\} */C */C) */C) & 305 & 1 \\
17 & (*/P.\{ox\} (*/B.\{ma\} */C */C) */S) & 253 & - \\
\rcg{}18 & (*/P.\{sr\} */C */S) & 249 & 1 \\
19 & (*/P.\{pa\} */C */S) & 232 & 1 \\
\rcg{}20 & (*/P.\{sc\} (*/B.\{ma\} */C */C) */C) & 215 & 1 \\
21 & (*/B.\{aa\} */C */C) & 202 & - \\
\rcg{}22 & (+/B.\{aa\} */C */C) & 202 & - \\
23 & (*/P.\{sx\} (*/B.\{ma\} */C */C) */S) & 176 & 3 \\
\rcg{}24 & (*/P.\{px\} (*/B.\{ma\} */C */C) */S) & 162 & 4 \\
25 & (*/P.\{sxr\} */C */S */R) & 161 & 3 \\
\rcg{}26 & (*/P.\{sor\} */C */C */R) & 158 & 1 \\
27 & (*/P.\{pr\} */C */R) & 150 & 1 \\
\rcg{}28 & (*/P.\{sox\} (*/B.\{ma\} */C */C) */C */S) & 147 & 1 \\
29 & (*/P.\{scx\} */C (*/B.\{ma\} */C */C) */S) & 121 & 5 \\
\rcg{}30 & (*/P.\{ox\} */C */R) & 120 & - \\
31 & (*/P.sr (*/B.\{ma\} */C */C) */R) & 117 & 1 \\
\rcg{}32 & (*/P.\{sxr\} */C (*/T */C) */R) & 115 & 3 \\
33 & (*/P.\{scx\} (*/B.\{ma\} */C */C) */C */S) & 111 & 1 \\
\rcg{}34 & (*/P.\{sox\} */C */C */R) & 109 & 1 \\
35 & (*/P.so (+/B.\{ma\} */C */C) */C) & 93 & 1 \\
\rcg{}36 & (*/P.\{pax\} */C */S */S) & 90 & 1 \\
37 & (*/P.\{pax\} */C (*/T */C) */S) & 86 & 1 \\
\rcg{}38 & (*/P.\{sc\} (*/B.\{mm\} */C */C) */C) & 84 & 1 \\
39 & (*/P.\{sc\} (+/B.\{mm\} */C */C) */C) & 84 & 1 \\
\rcg{}40 & (*/P.\{cx\} */C */S) & 81 & - \\
41 & (*/P.\{sxr\} */C */S */S) & 80 & 1 \\
\rcg{}42 & (*/P.\{sr\} (*/B.\{ma\} */C */C) */S) & 76 & 1 \\
43 & (*/P.\{sx\} (+/B.\{ma\} */C */C) */S) & 70 & - \\
\rcg{}44 & (*/P.\{sxr\} */C (*/T */C) */S) & 69 & 3 \\
45 & (*/P.\{sc\} (+/B.\{ma\} */C */C) */C) & 69 & 1 \\
\rcg{}46 & (*/P.\{sox\} (+/B.\{ma\} */C */C) */C */S) & 68 & 1 \\
47 & (*/P.\{scx\} */C */C */R) & 66 & 1 \\
\rcg{}48 & (*/P.\{sx\} */C */R) & 66 & - \\
49 & (*/P.\{ox\} (+/B.\{ma\} */C */C) */S) & 65 & - \\
\rcg{}50 & (*/P.\{or\} */C */R) & 65 & - \\
\bottomrule
\end{tabular}
}
\end{center}
\caption{50 most commons hyperedge patterns found in an SH extracted from a Wikipedia sample, excluding modifiers and conjunctions and restricting expansions to depth two. Only relations with 2 or 3 arguments are accepted, and the special builder \h{+/B} is explicitly considered.}
\label{tab:wikipedia-patterns}
\end{table*}

\end{document}